\documentclass[a4paper,11pt]{article}
\usepackage{jcappub}
\usepackage[T1]{fontenc} 

\usepackage{amsmath} 
\arxivnumber{2507.15436}

\title{\huge{Semi-empirical Framework of Supermassive Black Hole
Evolution: Highlighting a possible tension between Demographics and Gravitational Wave Background}}

\author[a,b,c,d,*]{Andrea Lapi,}
\author[e]{Francesco Shankar,}
\author[a,f]{Michele Bosi,}
\author[e]{Daniel Roberts,}
\author[g,e]{Hao Fu,}
\author[e]{Karthik M. Varadarajan,}
\author[h,*]{Lumen Boco}

\affiliation[a]{SISSA, Via Bonomea 265, 34136 Trieste, Italy}
\affiliation[b]{IFPU, Via Beirut 2, 34014 Trieste, Italy}
\affiliation[c]{INFN-Sezione di Trieste, via Valerio 2, 34127 Trieste, Italy}
\affiliation[d]{IRA-INAF, Via Gobetti 101, 40129 Bologna, Italy}
\affiliation[e]{Dept. of Physics and Astronomy, Univ. of Southampton, Highfield
SO17 1BJ, UK}
\affiliation[f]{Dept. of Physics, University of Trento, Via Sommarive 14, 38123 Povo (TN), Italy}
\affiliation[g]{Dept. of Physics, Fudan Univ., Shanghai 200438, PR China}
\affiliation[h]{Univ. Heidelberg, Albert-Ueberle-Str. 3, 69120 Heidelberg, Germany}

\affiliation[*]{corresponding authors}
\emailAdd{lapi@sissa.it,lumen.boco@uni-heidelberg.de}

\abstract{The evolution of the supermassive Black Hole (BH) population across cosmic times remains a central unresolved issue in modern astrophysics, due to the many noticeable uncertainties in the involved physical processes that span a huge range of spatial, temporal and energy scales. Here we tackle the problem via a semi-empirical approach with minimal assumptions and data-driven inputs. This is based on a continuity plus Smoluchowski equation framework that allows to unitarily describe the two primary modes of BH growth: gas accretion and binary mergers. Key quantities related to the latter processes are incorporated through educated parameterizations, and then constrained in a Bayesian setup from joint observational estimates of the local BH mass function, of the large-scale BH clustering, and of the nano-Hz stochastic gravitational wave (GW) background measured from Pulsar Timimg Array (PTA) experiments. We find that the BH accretion-related parameters are strongly dependent on the local BH mass function determination: higher normalizations and flatter high-mass slopes in the latter imply lower radiative efficiencies and mean Eddington ratios with a stronger redshift evolution. Additionally, the binary BH merger rate is estimated to be a fraction $\lesssim 10^{-1}$ of the galaxy merger rate derived from galaxy pairs counts by \texttt{JWST}, and constrained not to exceed the latter at $\gtrsim 2\sigma$. Relatedly, we highlight hints of a possible tension between current constraints on BH demographics and the interpretation of the nano-Hz GW background as predominantly caused by binary BH mergers. Specifically, we bound the latter's contribution to $\lesssim 30-50\%$ at $\sim 3\sigma$, suggesting that either systematics in the datasets considered here have been underestimated so far, or that additional astrophysical/cosmological sources are needed to explain the residual part of the signal measured by PTA experiments.}

\begin{document}
\maketitle
\flushbottom

\section{Introduction}\label{sec|intro}

Relic black holes (BHs) with masses $M_\bullet\sim 10^6-10^{10}\, M_\odot$ are found ubiquitously at the center of massive galaxies in the local Universe (e.g., \cite{Magorrian1998,Ferrarese2000,Gebhardt2000,Tremaine2002,Burke2025,Zou2025}). These monsters are thought to have grown mainly by gaseous accretion onto a disk surrounding the BH (e.g., \cite{LyndenBell1969,Shakura1973}) that energizes the spectacular broadband emission of active galactic nuclei (AGNs) over a broad range of bolometric luminosities $L_{\rm AGN}\sim 10^{42}-10^{48}$ erg s$^{-1}$. Such a paradigm has received a smoking gun confirmation by the Event Horizon Telescope \cite{EHT2019,EHT2022} via the imaging of the BH shadow caused by gravitational light bending and photon capture around the event horizon of M87 and Sgr A$^\star$. Nevertheless, the evolutionary history of the supermassive BH population is still an open hot issue of modern astrophysics, as testified by the many studies conducted to address it in the last decades via numerical simulations (e.g., \cite{Hirschmann2014,RosasGuevara2016,Volonteri2016,Springel2018,Weinberger2018,Dave2019,Ni2022,Chakraborty2023,Zou2024}), semi-analytic models (e.g., \cite{Kauffmann2000,Bower2006,Croton2006,Somerville2008,Dayal2019,Fontanot2020,Valiante2021,Ghodla2025}), and data-driven approaches (e.g., \cite{Hopkins2006,Shankar2013,Aversa2015,Tucci2017,Carraro2020,Zhang2023,Bosi2025}). 

In fact, some uncertainties still hinder a robust reconstruction of the growth history of supermassive
BHs, even on a population-averaged level. First of all, the initial conditions are poorly known.
The simplest picture envisaging light seeds with $M_\bullet \lesssim 10^{2}\, M_\odot$ of stellar origin
growing to the supermassive range by standard disk accretion has been challenged by the 
discovery of an increasing number of active BHs with masses $M_\bullet \gtrsim 10^9\, M_\odot$ already in place at appreciably high redshifts $z\gtrsim 7$, when the age of the universe was shorter than $0.8$ Gyr (e.g., \cite{Mortlock2011, Wu2015,Venemans2018,Banados2018,Reed2019,Matsuoka2019,Yang2021,Wang2021,Farina2022,Dodorico2023}; for a recent review see \cite{Fan2023}). This may require mechanisms able to rapidly produce  heavy BH seeds $M_\bullet \sim 10^{3-5}\, M_\odot$, thus reducing the time needed to attain the final masses (see \cite{Natarajan2014,Madau2014super,Mayer2019,Inayoshi2020,Boco2020,Volonteri2021,Pacucci2024}).

At the other end, the local demographics of the BH population still suffers from some systematics and possible biases. In fact, large samples of relic, quiescent BHs with precise mass determinations in the local Universe are not available. Thus the local BH mass function (i.e., the comoving number density of BHs as a function of mass) is usually estimated indirectly. A small sample with robust BH mass measurements is exploited to calibrate scaling relations of $M_\bullet$ with other host galaxy properties, like the stellar mass $M_\star$ or the stellar velocity dispersion $\sigma$ (e.g., \cite{Kormendy2013,McConnell2013,Reines2015,Davis2018,Sahu2019,Zhu2021}). Such scaling relations are then employed to convert the statistics of a given galaxy property (e.g., the stellar mass function or the velocity dispersion function, determined on much larger samples of galaxies) into the BH mass function. However, this method is subject to systematics uncertainties that may drive the observed scaling relations away from the intrinsic ones \cite{Bernardi2007,Shankar2016,Shankar2019,Shankar2020}, hence potentially producing biased estimates of the local BH mass function.

In addition, the intermediate stages of the BH evolution are also somewhat debated. For example, the accretion rates $\dot M_\bullet \propto L_{\rm AGN}/\epsilon$ and radiative efficiencies $\epsilon$ for the population of active BHs are loosely constrained, especially toward high-redshift \cite{Soltan1982,Salucci1999,Davis2011,Raimundo2012,Trakhtenbrot2017,Aird2022,Laloux2024}. The same applies to the distribution of Eddington ratio $\lambda\propto L_{\rm AGN}/M_\bullet$ (e.g., \cite{Kollmeier2006,Vestergaard2009,Willott2010,Kelly2013,Farina2022,Wu2022}) or of its proxy $\lambda_\star\propto L_X/M_\star$ involving the X-ray luminosity $L_X$ and host stellar mass $M_\star$ (e.g., \cite{Bongiorno2016,Georgakakis2017,Aird2018,Carraro2022}), which are only vaguely assessed for $z\gtrsim 4$. Even the redshift-dependent AGN luminosity function estimated via multi-wavelength surveys (e.g., \cite{Hopkins2007LF,Shen2020}) may suffer at $z\gtrsim 6$ from residual systematics on the adopted bolometric corrections and obscured fractions especially at the faint end (e.g., \cite{Lusso2012,Duras2020,Ueda2014,Buchner2015,Ananna2019,Laloux2023,Andonie2024}). The overall picture has been made even more complex by the recent observations with \texttt{JWST}, which have identified at $z\gtrsim 4.5$ an abundant population of overmassive, X-ray under-luminous BHs (e.g., \cite{Ubler2023,Furtak2023,Harikane2023,Kokorev2023,Stone2024}); several of these are located in the so called `little red dots' sources (see \cite{Kocevski2023,Matthee2024,He2024,Lai2024,Kokorev2024}), whose number density at $z\gtrsim 6$ may be up to two orders of magnitude higher than that of UV/optically selected AGNs \cite{Kokorev2024,Akins2025}.

Finally, the role of BH mergers, which constitute another relevant channel for the growth of supermassive BHs, is still under scrutiny. Binary BH merger events are thought to follow (maybe with some delay) the possible aggregation of the host galaxies, common in overdense environments especially toward low redshifts. However, so far such events have remained elusive and difficult to be corroborated, even indirectly, by observations of dual AGNs (e.g., \cite{Comerford2013,Silverman2020,Pfeifle2025}). Recently, the situation may have changed with the possible detection of the stochastic gravitational wave (GW) background in the nano-Hz frequency range by pulsar timing array (PTA) experiments \cite{Agazie2023,EPTA2024,Reardon2023,Xu2023}. Although the statistical significance of the signal does not attain the $5\sigma$ threshold of confidence, and superimposed contributions from different physical mechanisms may prevent a clear-cut interpretation, such data offer a new observational handle on the BH growth process via binary mergers, as discussed in the literature with analytic approaches \cite{Wyithe2003,Sesana2008,Kulier2015,Sesana2016,Dascoli2018,Chen2019,Simon2023,SatoPolito2024,SatoPolito2025} and numerical simulations \cite{Salcido2016,Kelley2017,Siwek2020,Chen2025}.

To address the numerous facets and issues of such a complex landscape, three main complementary approaches are possible and have been largely exploited in the last decades: hydrodynamical simulations (for a review, see \cite{Naab2017}), semi-analytic models (for a review, see
\cite{Somerville2015rev}), and semi-empirical frameworks (for a review, see \cite{Lapi2025}). Hydrodynamical simulations allow to address the galaxy and BH evolution process in fine detail, as they can treat simultaneously dark matter, gas and stars. However, despite the recent increase in resolution and speed (mainly thanks to emulators and 'genetic' codes), many of the relevant physical processes still constitute sub-grid physics, while a detailed exploration of the parameter space is often limited by rather long computational times. 
Semi-analytic models are based on dark matter merger trees extracted from $N-$body simulations (or generated via Monte Carlo procedures gauged on them), while the physics inside
dark halos is modeled via parametric expressions set on (mainly) local observables. These models are less computationally expensive than hydrodynamical simulations and enable to test  different prescriptions more easily, though the considerable number of fudge parameters describing the underlying physics can lead to degenerate solutions and restrict somewhat the predictive power, especially toward high redshift (e.g., \cite{Seeyave2025}). Concerning the evolution of the (super)massive BH population, ab-initio models struggle somewhat in reproducing coherently the wealth of observational data and in elucidating the different role of the main physical mechanisms at work. This is mainly because the problem requires to self-consistently model many diverse processes cooperating on a huge range of temporal, spatial, and energy scales. Such a task goes by far beyond our present physical understanding and computational capabilities, and in any case tends to produce outcomes strongly contingent on the chosen modeling, sub-grid recipes and spatial/temporal resolution (see \cite{Donnari2021,Habouzit2022,Crain2023}).

The situation calls for a reverse-engineering approach that inverts the traditional methodology: rather than testing ab-initio model predictions against observations, one can construct semi-empirical frameworks that extract physical insights directly from data with minimal theoretical assumptions. While these frameworks may not capture the full theoretical richness of ab-initio models or provide complete physical narratives, their strength lies in delivering robust, model-independent constraints on key observables and providing clear answers to specific, well-posed questions.
Moreover, by empirically linking different observables, they can check for possible inconsistencies or tensions among distinct datasets.
Such data-driven models serve not to replace more sophisticated semi-analytic and numerical approaches, but are meant as complementary tools to provide independent constraints that can break existing degeneracies and guide theoretical developments toward more physically-motivated parameterizations.

In this work we approach the issue of (super)massive BH evolution via a semi-empirical framework, which is anchored by construction to the most reliable data available so far, takes into account  observational/systematic uncertainties by transparently highlighting their impact on the results, and aims to minimize theoretical assumptions and associated parameterizations.  Specifically, we employ a continuity plus Smoluchowski equation framework to unitarily describe the two main modes of growth for the supermassive BH population, namely gas accretion and binary mergers. We educatedly parameterize the main uncertainties associated to each BH growth channel, and infer joint constraints on these in a Bayesian setup from estimates of the local BH demographics, clustering, and measurements of the nano-Hz stochastic GW background.  The plan of the paper is as follows: in Section 2 we describe the ingredients and methodology underlying our semi-empirical framework; in Section 3 we present our data analysis and discuss our results; in Section 4 we summarize our findings. Throughout this work we adopt the standard $\Lambda$CDM cosmology with parameters from the \texttt{Planck} mission \cite{Planck2020}.

\section{Methodology}\label{sec|theory}

\begin{figure}[t!]
    \centering
  \rotatebox{90}{
    \begin{minipage}{1.5\textwidth}
      \includegraphics[width=1\linewidth]{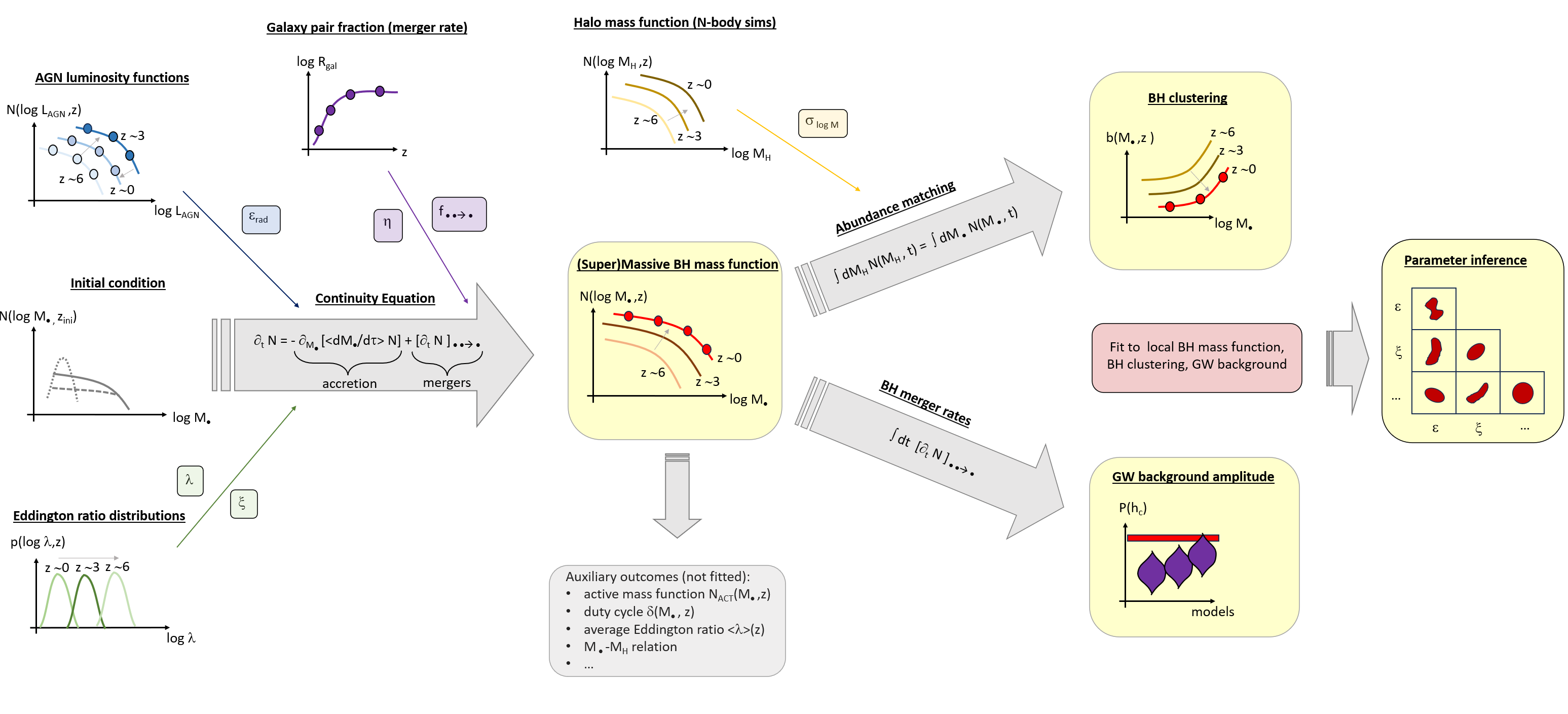}
      \caption{Schematics showing the basic methodology underlying our semi-empirical framework (see Section \ref{sec|theory} for details).}\label{fig|schematics}
    \end{minipage}}
\end{figure}

In this Section we first provide a qualitative overview of our data-driven methodology, which is also schematically illustrated in Figure \ref{fig|schematics}, while formal details and references can be found in the following subsections.

\begin{itemize}

\item The core tool of the method is constituted by the continuity equation. This provides an effective way of deriving the statistical evolution of the (super)massive BH population from data-driven inputs. Specifically, the continuity equation is a partial differential equation which evolves timestep by timestep the BH mass function across cosmic time, taking into account the averaged growth of the BH population due to gas accretion and mergers.

\item The component related to gas accretion is described by a drift term in the continuity equation. It requires two \emph{inputs}: the observed redshift-dependent bolometric AGN luminosity function $N(\log L_{\rm AGN},t)$, which provides the statistics of active BHs emitting at a given luminosity $L_{\rm AGN}$; a parameterized form of the Eddington ratio distribution $P(\log \lambda)$, which basically informs on how BHs of different masses are contributing to the different luminosity bins. The \emph{parameters} involved are three: two ($\log \lambda_0$ and $\xi$) describe the local normalization and redshift evolution for the mean of the Eddington ratio distribution; the other one is the radiative efficiency $\epsilon$, which allows to convert AGN luminosity into BH accretion rates, eventually needed to compute BH masses. 

\item The component related to mergers is described by a Smoluchowski coagulation (kind of diffusion) term in the continuity equation. It requires one \emph{input} and two additional \emph{parameters}: in the reasonable assumption that BH mergers follow galaxy mergers, one parameter $f_{\bullet\bullet\rightarrow \bullet}$ is a rescaling factor for the the redshift-dependent normalization of the BH merger rate with respect to the galaxy merger rate as estimated from galaxy pair counts, which constitutes the observational input (but we will explore variants where the BH merger rate is maximized by adopting $f_{\bullet\bullet\rightarrow \bullet}\approx 1$); another parameter $\eta$ specifies a possible additional dependence of the BH binary merger rate on the BH mass.

\item The \emph{direct output} of the continuity equation is a unique solution for the amplitude, shape and evolution of the BH mass function $N(\log M_\bullet,t)$ as built by accretion and mergers. The BH mass function at $z\approx 0$ can be directly compared with independent observational determinations from local galaxy statistics and BH-to-galaxy scaling relations. 

\item A first \emph{derived outcome} is the BH clustering. We compute the mean relation between BH mass and host dark matter halo mass via abundance matching of the BH mass function yielded by the continuity equation and the halo mass function $N(\log M_{\rm H},t)$ as provided by $N-$body simulations in the standard $\Lambda$CDM cosmology. We then use this relation to convert halo bias into BH bias (i.e., a clustering signal). This procedure requires an additional \emph{parameter}, which is the scatter $\sigma_{\log M_{\rm H}}$ in the relationship between BH and host halo mass. The BH bias so obtained can be directly compared with observational estimates of BH clustering.

\item Another \emph{derived outcome} is the stochastic GW background from BH mergers in the nano-Hz frequency range. This can be straightforwardly computed from the outputs of the continuity equation and in particular from the redshift-dependent BH mass function and binary BH merger rate (related to the Smoluchowski coagulation term). The resulting GW background amplitude can then be directly compared with the current observational estimates from PTA experiments.

\item The aforementioned parameters of our semi-empirical framework (namely, the radiative efficiency of BH accretion, the  mean of the Eddington ratio distribution and its redshift evolution, the normalization and mass dependence of the BH merger rate, and the scatter in the BH vs. halo mass relation) are \emph{inferred} by comparing the direct output and the derived outcomes with independently measured datasets within a fully Bayesian setup. This approach enables a robust quantification of parameter degeneracies and statistical uncertainties. A key strength of our framework is its ability to jointly constrain the evolution of the supermassive BH population by simultaneously fitting observational data on BH demographics, large-scale clustering, and the GW background. 
We will discuss below how such a combination of independent observables enables us to highlight a possible tension between the amount of GW background due to binary BH merger and the measured signal from PTA experiments, which could instead be missed by BH merger models that attempt to only fit the GW background ignoring all the other demographic indicators. 

\end{itemize}

In summary, the continuity equation is a flexible framework that by integrating a few observational inputs (e.g., AGN luminosity functions) yields a clear direct output (e.g., BH mass function) and some derived outcomes (e.g., BH clustering, PTA signal) via minimal physical assumptions (e.g., BHs grow by gas accretion and binary mergers) and a limited number of parameters (six in total, see above). The direct output and derived outcomes are then compared with independent datasets to obtain a best estimate of the parameters, which ultimately yields insights into the physical evolution of the BH population. Such a data-driven approach bypasses the complex and uncertain modeling of the baryonic physics associated to galaxies within dark matter halos via semi-analytic recipes or hydrodynamical simulations. Moreover, since the number of observables and parameters are comparable, it is not prone to large degeneracies but only affected by possible systematics. All in all, our semi-empirical framework can robustly assess the relative contributions of accretion and mergers to the evolutionary history of the supermassive BH population, whilst constraining the fraction of the GW background that can plausibly be attributed to supermassive BH mergers. 

\subsection{The continuity equation}

As introduced above, a very effective approach to study the statistical evolution of the supermassive BH population in a data-driven way relies on the continuity equation \cite{Cavaliere1971,Small1992, Marconi2004,Shankar2004,Yu2008,Cao2010,Shankar2013,Aversa2015,Tucci2017,Sicilia2022}. This method allows to self-consistently trace the cosmic evolution of the BH mass function and the associated fraction of active BHs, building on the amount of mass accreted at each cosmic epoch. The latter is extracted from the input AGN luminosity function, from the adopted relation between AGN luminosity and BH mass as specified by the Eddington ratio distribution of active BHs, and from the assumed radiative (plus kinetic) efficiency.  Moreover, being based on a linear differential equation, this approach allows to transparently disentangle the relative roles of accretion and binary mergers as complementary growth channels of the BH population at each cosmic time.

Let $N(M_\bullet,t)$ be the BH mass function, i.e. number density (per comoving volume) of BHs with mass $M_\bullet$ at cosmic time $t$ or at the corresponding redshift $z$. In the reasonable hypothesis that any individual BH can change mass because of gas accretion and/or of binary merger with a companion, then the evolution of the mass function can be described via the continuity equation:
\begin{equation}\label{eq|basic}
\partial_t\, N(M_\bullet,t) = -\partial_{M_\bullet} [\langle\dot M_\bullet\rangle(M_\bullet,t)\, N(M_\bullet,t)] + [\partial_t N(M_\bullet,t)]_{\bullet\bullet\rightarrow \bullet} \; .
\end{equation}
The first addendum on the right hand side is a `drift' term describing growth by gas accretion via a mean accretion rate $\langle\dot M_\bullet\rangle$ averaged over the BH population; the underlying rationale is that, although individual BHs may turn on and off, the statistical evolution depends only on the mean accretion rate as a function of mass and time. The second addendum $[\partial_t N]_{\bullet\bullet\rightarrow \bullet}$ is a `coagulation' term accounting for binary mergers between BHs; note that mergers do not add net mass to the BH population but can alter the mass function by redistribution of the masses, from the low to the high mass end\footnote{Note, however, that numerically solving the continuity equation on a finite grid of BH masses requires to assume that a constant flux of sub-resolution masses filtrates through the lowest mass boundary, see \cite{Shankar2009}.}. We now turn to describe these two ingredients in some more detail.

\subsection{Growth by accretion}\label{sec|accretion}

\begin{figure}[t!]
    \centering
    \includegraphics[width=1.\textwidth]{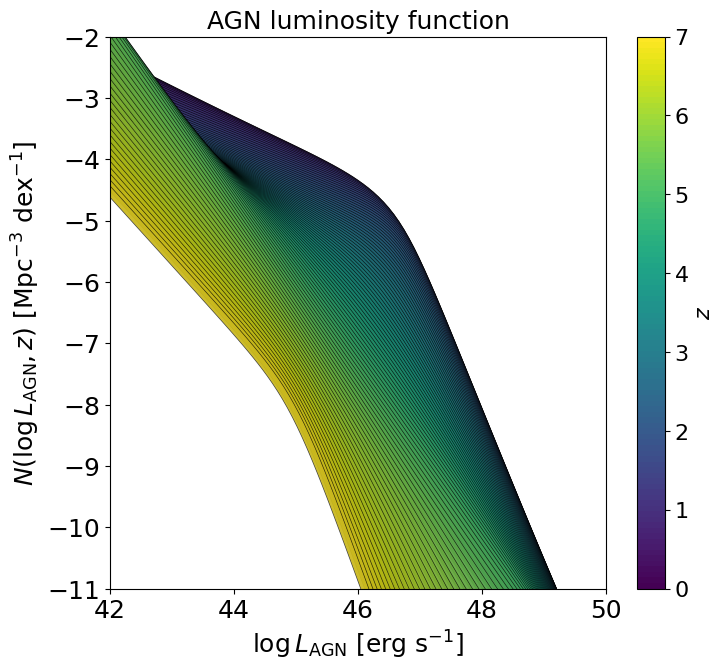}
    \caption{The AGN bolometric luminosity functions derived from multi-wavelength data by \cite{Shen2020} (we use the fit labeled 'A' in their Table 4; see also Section \ref{sec|accretion}) as a function of redshift (color-coded), which is the basic input in our semi-empirical framework.}
    \label{fig|AGNLF}
\end{figure}

To compute the growth by accretion we rely on one crucial observational input, constituted by the redshift-dependent AGN bolometric luminosity function $N(\log L_{\rm AGN},t)$ derived from multi-wavelength data in optical, UV, IR and X-ray bands by \cite{Shen2020}. Specifically, we use the analytic rendition by \cite{Shen2020} labeled 'A' in their Table 4. This is expressed as
\begin{equation}\label{eq|shen}
N(\log L_{\rm AGN}) = \frac{\phi_\star}{(L_{\rm AGN}/L_\star)^{\gamma_1}+(L_{\rm AGN}/L_\star)^{\gamma_2}}
\end{equation}
in terms of the parameters $\{\phi_\star(z),\gamma_1(z),\gamma_2(z),L_\star(z)\}$. The latter depend on redshift $z$ according to the expressions
\begin{equation}
\begin{aligned}
& \log \phi_\star\, [{\rm Mpc}^{-3}\, {\rm dex}^{-1}]  = d_0+d_1\,(1+z)\\
\\
& \gamma_1 =a_0+a_1\,(1+z)+a_2\,[2\,(1+z)^2-1]\\
\\
& \gamma_2 = \frac{2\,b_0}{[(1+z)/3]^{b_1}+[(1+z)/3]^{b_2}}\\
\\
& \log L_\star\,[L_\odot] = \frac{2\,c_0}{[(1+z)/3]^{c_1}+[(1+z)/3]^{c_2}}\\
\end{aligned}
\end{equation}
where $\{a_0,a_1,a_2\}=\{0.8569,-0.2614,0.02\}$, $\{b_0,b_1,b_2\}=\{2.5375,-1.0425,1.1201\}$, $\{c_0,c_1,c_2\}=\{13.0088,-0.5759,0.4554\}$, $\{d_0,d_1\}=\{-3.5426,-0.3936\}$. The resulting bolometric AGN luminosity functions are plotted as a function of redshift over the range $z\sim 0-7$ (color-coded) in Figure \ref{fig|AGNLF}.

We exploit the AGN luminosity function to compute the accretion rate $\langle \dot M_\bullet\rangle$ averaged over the population of active BH, which enters in the drift term of the continuity Equation (\ref{eq|basic}). To make the computation explicit, we need some preliminary definitions and steps. To start with, it is convenient to relate the AGN luminosity $L_{\rm AGN}$ to the BH mass $M_\bullet$. One defines the Eddington ratio $\lambda\equiv L_{\rm AGN}/L_{\rm Edd}$, which is the ratio between the actual AGN luminosity and the Eddington luminosity. The Eddington luminosity $L_{\rm Edd} \equiv  4\pi G \mu_e m_p\,c\, M_\bullet/\sigma_{\rm T} \approx 1.3 \times 10^{38}\, (M_\bullet/M_\odot)$ erg s$^{-1}$ is the limiting value for which the continuum radiation force emitted by the accretion disk around the BH balances gravity in isotropic conditions; in compact form one can write $L_{\rm Edd} = M_\bullet c^2/t_{\rm Edd}$ in terms of the Eddington timescale $t_{\rm Edd} = c\,\sigma_{\rm T}/4\pi G\,\mu_e m_p\approx 0.45$ Gyr, only dependent on fundamental constants\footnote{In this definition, $G$ is the gravitational constant, $\mu_e$ is the mean molecular weight per electron, $m_p$ is the proton mass, $c$ is the speed of light, and $\sigma_{\rm T}$ is the Thomson cross section.}. Provided these definitions, $L_{\rm AGN} = \lambda\, M_\bullet\,c^2/t_{\rm Edd}$ is just a rewriting that by itself 
does not require any assumption and does not add any relevant information unless one specifies how the Eddington ratio (or the AGN luminosity to BH mass ratio) varies during the lifetime of an individual AGN. However, this rewriting is helpful in view of a statistical description like the one subtended by the continuity equation, because as shown below the demographics (i.e., the mass function) of \emph{active} BHs can be fully characterized by the observed AGN luminosity function and by the redshift-dependent distribution of Eddington ratios.

In fact, as suggested by observations in the local and distant Universe \cite{Kollmeier2006,Willott2010,Kelly2013,Farina2022,Wu2022,He2024,Lai2024,Burke2025,Zou2025}, active BHs features a distribution of Eddington ratios  with a lognormal shape\footnote{We also try to adopt a Schechter-like or a lognormal$+$powerlaw shape, which have been sometime considered in the literature as inspired from theoretical arguments or empirical analysis of type-2 AGNs at relatively low $z$ (e.g., \cite{Hopkins2006,Kauffmann2009,Yu2008,Shankar2013,Aversa2015,Tucci2017}. This requires to add at least one parameter (e.g., powerlaw behavior at small BH masses and its redshift evolution), which however we find to be largely unconstrained given the datasets considered in this work. Thus we prefer to use the pure log-normal expression involving the minimal set of parameters. In any case, we checked that qualitatively our results concerning the buildup of the BH mass function, the role of BH mergers and their contribution to the GW background are unaffected by the choice of the Eddington ratio distribution shape; minor quantitative alterations pertain only to the evolution of the BH-accretion parameters.}
\begin{equation}\label{eq|plambda}
p(\log\lambda) = \frac{1}{\sqrt{2\pi}\sigma_{\log\lambda}}\,e^{-(\log \lambda-\langle\log\lambda\rangle)^2/2\sigma_{\log\lambda}^2}\; ,
\end{equation}
with dispersion $\sigma_{\log\lambda}\approx 0.35$ dex and with a redshift dependent mean. For the latter we adopt the scaling
\begin{equation}\label{eq|lambda}
\langle\log\lambda\rangle(z)=\log \lambda_0+ \xi\,\log(1+z) \; ,
\end{equation}
where $\lambda_0$ and $\xi$ are fitting parameters to be determined. We also tried to add a dependence on BH mass, but we found it to be unconstrained with the current data on BH demographics, so we decided to drop it from our analysis. 

Then the active BH mass function is linked to the observed AGN luminosity function and to the Eddington ratio distribution by the convolution
\begin{equation}\label{eq|conv}
N(\log L_{\rm AGN},t) = \int{\rm d}\log \lambda\; p(\log\lambda)\, N_{\rm AGN}(\log M_\bullet,t)_{\big|M_\bullet=L_{\rm AGN}\, t_{\rm Edd}/\lambda c^2}\; ;
\end{equation}
this equation may be solved parametrically by assuming a standard double power-law shape\footnote{Although assuming a double power-law shape for the active BH mass function is standard practice, it is worth making a disclaimer here. There is no question that the AGN luminosity function is well described by a double power-law behavior. Since we have assumed a lognormal shape for the Eddington ratio distribution to reflect the latest observational determinations, we are naturally led to adopt a double power-law shape for the active mass function, otherwise the convolution Equation (\ref{eq|convo}) would not produce the correct shape to fit the luminosity functions. On the other hand, we point out that such an assumption has the direct consequence of yielding a power-law shape in the relic BH mass function at the high-mass end. Despite this is somewhat supported by the data (see Section \ref{sec|results}), the presence of an exponential falloff is still not excluded. Forcing the accretion model to produce it  could be achieved by employing a Schechter fit for the active mass function. However, this solution would reduce the high-mass end of the predicted BH mass function at all redshifts, so lowering the BH merger rates and exacerbating the tension with the GW background (see Section \ref{sec|results}).} \cite{Schulze2010,Cao2010,Shankar2013} for the active mass function $N_{\rm AGN}=N_0/[(M_\bullet/M_{\bullet,0})^{\omega_1}+(M_\bullet/M_{\bullet,0})^{\omega_2}]$ and then by fitting for the best parameters $(N_0,M_{\bullet,0},\omega_1,\omega_2)(t)$ at any redshift to reproduce the AGN luminosity function. Note that another way to proceed will be to physically model the Eddington ratio distribution by assuming an average light-curve shape for active BHs, as it has been pursued in \cite{Hopkins2006,Yu2008,Aversa2015}; however, in the spirit of a data-driven model here we do not take such a route.

To proceed further, we need the BH accretion rate associated to a given AGN luminosity. This is provided by $\dot M_\bullet = (1-\epsilon-\epsilon_{\rm kin})\,L_{\rm AGN}/\epsilon\, c^2$ where $\epsilon$ and $\epsilon_{\rm kin}$ are the radiative and kinetic (jets and/or winds) efficiency, respectively\footnote{The radiative efficiency $\epsilon$ is conventionally defined with respect to the large-scale mass inflow rate $\dot M_{\rm inflow}$ so that $L_{\rm AGN} = \epsilon\, \dot M_{\rm inflow} c^2$, but the actual BH growth rate $\dot M_\bullet = (1 - \epsilon-\epsilon_{\rm kin})\,\dot M_{\rm infl}$ is smaller because of radiative and kinetic losses. Putting together these expressions yields $\dot M_\bullet = (1-\epsilon-\epsilon_{\rm kin})\,L_{\rm AGN}/\epsilon\, c^2$.}. Using the previous rewriting of $L_{\rm AGN}$ in terms of BH mass $M_\bullet$, this can also be expressed as $\dot M_\bullet = M_\bullet/\tau_{\rm ef}$ in terms of the $e-$folding time $\tau_{\rm ef}\equiv \epsilon\, t_{\rm Edd}/\lambda\,(1-\epsilon-\epsilon_{\rm kin})$, which would be the theoretical timescale required for the BH to exponentially grow $M_\bullet\propto e^{\tau/\tau_{\rm ef}}$ by a factor $e\approx 2.7$ if $\lambda$ were constant in time. For an average efficiency $\epsilon\sim 0.1$ and $\lambda\sim 1$ it takes on values $\tau_{\rm ef}\approx 5\times 10^7$ yr, which is often quoted in textbooks as the typical growth timescale for a BH. However, we stress that such a timescale does not directly correspond to the duty cycle or visibility timescale of an AGN, first because for individual AGNs $\lambda$ is not constant along the light-curve, second because the emission could be stochastic or strongly intermittent, and finally because the visibility time depends on environmental conditions (e.g., amount of obscuration), on the observational band, and on selection effects (see \cite{AlonsoTetilla2025}). For example, literature estimates of the visibility time amount to $\lesssim$ a few $10^7$ yr in the optical band for unobscured quasars, while they can be as high as $\gtrsim 10^8$ yr in the hard X-ray band for heavily obscured AGNs \cite{Shankar2004,Lapi2006}.

Some discussion on the radiative efficiency $\epsilon$ is in order here. For a standard thin disk the radiative efficiency may range from $\epsilon\sim 0.06$ for a non-rotating to $\epsilon\sim 0.42$ for a maximally spinning Kerr BH \cite{Thorne1974,Davis2011}. However, both for high accretion rates with $\lambda\gtrsim $ a few and for very low accretion rates with $\lambda\lesssim 0.01$, the radiative efficiency tends to decrease due to the onset of an advection dominated regime. In the former case, mainly applying at high redshift, the inflowing gas has a huge density and provides a very large optical depth to trap the radiation and advect it back into the BH \cite{Abramowicz1988,Watarai2000,Madau2014}. In the latter case, mainly applying to low redshift and in the re-activations of dormant BHs, the inflowing gas density is so low to hinder cooling within an accretion timescale, and to store the viscous energy as a thermal component that is eventually advected into the hole \cite{Narayan1995,Blandford2004,Merloni2008}. We have devised a handy $\lambda-$dependent expression for the radiative efficiency as a function of the Eddington ratio that describes these various regimes:
\begin{equation}\label{eq|epsilon}
\epsilon = \epsilon_{\rm thin}\, \Bigg[\frac{\lambda/2}{e^{\lambda/2}-1}-e^{-2\,(\lambda/\lambda_c)^s}\Bigg]\;,
\end{equation}
with $\epsilon_{\rm thin}$ a fitting parameter representing the efficiency of standard disk accretion; the quantities $\lambda_c\approx 0.01$ and $s\approx 1$ determine the behavior in the low accretion regime as suggested by the aforementioned studies, and are held fixed. The above detailed expression is adopted for the sake of completeness, but we anticipate that the advection dominated regimes will turn out to be marginally relevant for the evolution of the BH mass function. This is because the Eddington ratio distribution parameters inferred from our analysis will imply that the thin disk regime is the most typical accretion mode, with mildly super-Eddington values limited to high redshift $z\gtrsim 4-6$ and considerably sub-Eddington ones applying only to low $z\lesssim 1$ and large BH masses. As for the kinetic efficiency due to winds and jets, we assume $\epsilon_{\rm kin}\approx 0.15\,\epsilon_{\rm thin}$ as suggested by \cite{Shankar2008}; we anticipate that given the average values $\epsilon_{\rm thin}\lesssim 0.1$ inferred from our analysis, this position implies rather limited kinetic efficiencies. However, note that a few literature studies \cite{Ghisellini2013,Zubovas2018} claim rather higher $\epsilon_{\rm kin}\lesssim 0.2$, though not applying to all the BH population; in any case, we will explore such a possibility in Appendix \ref{app|tests}. Given the above, the only free parameter of our model specifying the relation between luminosity and accretion rate is the thin-disk radiative efficiency $\epsilon_{\rm thin}$. 

We have now all the ingredients to compute the mean accretion rate $\langle\dot M_\bullet\rangle$ entering Equation (\ref{eq|basic}). This is given by
\begin{equation}\label{eq|convo}
\langle\dot M_\bullet\rangle(M_\bullet,t) = \frac{M_\bullet}{t_{\rm Edd}}\,\int{\rm d}\log \lambda\; p(\log\lambda)\, \lambda\, \frac{1-\epsilon-\epsilon_{\rm kin}}{\epsilon}\,\frac{N_{\rm AGN}(\log M_\bullet,t)}{N(\log M_\bullet,t)}\; ,
\end{equation}
where $p(\log \lambda)$ is the Eddington ratio distribution from Equation (\ref{eq|plambda}), $N_{\rm AGN}(\log M_\bullet,t)$ is the active BH mass function from Equation (\ref{eq|conv}) and $\epsilon$ is the $\lambda-$dependent radiative efficiency from Equation (\ref{eq|epsilon}). To better catch the meaning of this expression, note that the quantity $\lambda\,(1-\epsilon-\epsilon_{\rm kin})\,M_\bullet/\epsilon\, t_{\rm Edd} = M_\bullet/\tau_{\rm ef}=\dot M_\bullet$ is just the accretion rate $\dot M_\bullet$ of an individual BH, and
the ratio $N_{\rm AGN}(\log M_\bullet,t)/ N(\log M_\bullet,t)\equiv \langle\delta\rangle(M_\bullet,t)$ between the mass function of active and overall BH population defines the so called
mean duty cycle $\langle\delta\rangle (M_\bullet,t)$. Thus the above expression can be also written in compact form as
$\langle\dot M_\bullet\rangle = \int{\rm d}\log\lambda\; p(\log \lambda) \dot M_\bullet \langle\delta\rangle$, highlighting that the underlying rationale is just to weight the accretion rate of a BH by the duty cycle and to average it over the Eddington ratio distribution.

Note that in the continuity equation only the combination $\langle \dot M_\bullet\rangle N(\log M_\bullet) = \int{\rm d}\log\lambda\,$ $p(\log\lambda)\,\dot M_\bullet\, N_{\rm AGN}$ enters the drift term describing accretion, and this depends just on the Eddington ratio distribution and the active BH mass function, not on the duty cycle. In fact, in our semi-empirical framework the duty cycle $\langle\delta\rangle(\log M_\bullet,z)$ is not modeled ab-initio and does not serve as an input (apart possibly for the initial condition, see Section \ref{sec|demographics}), but it is a derived outcome that can be computed after solving the continuity equation for the total BH mass function. However, a detailed comparison of the duty cycle so obtained with observations is not straightforward. First, because observational estimates of the duty cycle as a function of BH mass are appreciably uncertain, especially toward intermediate to high redshift. Second, because the estimates of the duty cycle depend on the observational band and on selection effects, like the luminosity threshold for an AGN to be considered as active. Nevertheless, we anticipate that in Section \ref{sec|results} our semi-empirical framework will be shown to produce local mean duty cycles in qualitative agreement with current observations.

It is worth stressing that for pure accretion (no mergers), constant radiative efficiency $\epsilon$ (no kinetic output) and Dirac-$\delta$ Eddington ratio distribution centered on a mean value $\langle\lambda\rangle$ independent of redshift, the duty cycle is just the ratio of the AGN luminosity and total BH mass function, so that the continuity equation can be readily integrated to yield
\begin{equation}\label{eq|easycont}
N(M_\bullet,t) = - \frac{(1-\epsilon)\, \langle\lambda\rangle^2\, c^2}{\epsilon\, t_{\rm Edd}^2\, \ln(10)}\,\int{\rm d}t\,\Bigg[\partial_{L_{\rm AGN}}\,N(\log L_{\rm AGN},t)\Bigg]_{\big|L_{\rm AGN}=\langle\lambda\rangle\, M_\bullet\, c^2/t_{\rm Edd}}\;.
\end{equation}
which recovers the simplest solution pioneered by \cite{Marconi2004,Shankar2004}. 

\subsection{Growth by mergers}\label{sec|mergers}

The most general framework\footnote{In the literature, the evolution of the BH mass function has been approached with a Fokker-Planck (diffusion-like) equation \cite{Yu2002,Vittorini2005,Shankar2013} but this is only an approximation. In fact, it can be shown (for details, see \cite{Leyvraz2003}) that the Smoluchowski equation reduces to a Fokker-Planck equation when the mass distribution is sharply peaked around a mean value and the system is in a regime where small changes in mass dominate. These conditions, however, are not guaranteed to hold during the overall evolution of the BH population. Moreover, the Fokker-Planck approximation is not appropriate to distinguish between the production and destruction terms due to binary mergers; an unambiguous assessment of the former is instead required to estimate the GW background.} to describe the evolution of the mass distribution $N(M_\bullet,t)$ for a population of BHs under binary aggregations is provided by the Smoluchowski coagulation equation \cite{Smoluchowski1916,Cavaliere1992,Benson2005,Erickcek2006,Ellis2024,SatoPolito2025}:
\begin{equation}\label{eq|Smolu}
\begin{aligned}
[\partial_t N(M_\bullet,t)]_{\bullet\bullet\rightarrow \bullet} & =  \frac{1}{2}\, \int_0^{M_\bullet}{\rm d}M_\bullet'\, K(M_\bullet',M_\bullet-M_\bullet',t)\, N(M_\bullet',t)\, N(M_\bullet-M_\bullet',t) +\\
&\\
& - N(M_\bullet,t)\, \int_0^\infty{\rm d}M_\bullet'\, K(M_\bullet,M_\bullet',t)\, N(M_\bullet',t)\;,
\end{aligned}
\end{equation}
where $K(M_\bullet,M_\bullet',t)$ is a time-dependent coagulation kernel, which has dimension of inverse of a time (i.e., a merging rate); the kernel is routinely taken to be 
symmetric in its mass dependencies, i.e. $K(M_\bullet,M_\bullet',t)=K(M_\bullet',M_\bullet,t)$. 
The first integral on the right hand side of the above equation is a production term which refers to a merging event between masses $M_\bullet'$ and $M_\bullet-M_\bullet'$ to produce the mass $M_\bullet$. The factor $1/2$ avoids over-counting since the labels of the two merging masses may be interchanged
given the symmetric nature of the kernel; it could be removed by changing the upper limit of integration to $M_\bullet/2$. The second addendum is a destruction term which accounts for merging between the mass $M_\bullet$ and a companion with mass $M_\bullet'$. 

We stress that the Smoluchowski equation is only appropriate when the population evolution is dominated by binary aggregations, and instead should not be employed when the probability of multiple interactions is appreciable. For example, it is not very apt to describe the merging history of dark matter halos at high redshift (or during any fast accretion phase), since the kernels inferred from the excursion set formalism or extracted from numerical simulations imply that the average number of halo mergers is often larger than two (e.g., \cite{Benson2005,Neistein2008}). However, for supermassive BHs the binary merger assumption is more reasonable. In fact, the issue has been somewhat investigated via aimed dynamical simulations \cite{AmaroSeoane2023,Koehn2023,Sayeb2024}. These show that only a minority around $10-20\%$ of binary supermassive BHs can form a triple system by capturing another BH intruder. Moreover, only a fraction $\lesssim 5\%$ actually form a triple systems at parsec scale experiencing chaotic three-body dynamical interaction; in the remaining majority of the cases two of the three BHs merge first and the third is ejected or left orbiting at large separation, so that the the merger is still binary. Thus despite merging triple BH systems could be interesting for future GW detector like \texttt{LISA} since they are expected to feature some peculiar signatures, their overall number is very small, and their contribution to the stochastic GW background irrelevant, compared to the standard binary merger channel.

If the mass function evolves under pure coagulation (i.e., no accretion, hence $\partial_t N\equiv [\partial_t N]_{\bullet\bullet\rightarrow \bullet}$) and if the kernel does not rise too steeply as a function of the mass, then by simple algebraic manipulation of Equation (\ref{eq|Smolu}) one can write a symmetrized equation for the moments of the mass function in the form
\begin{equation}\label{eq|Smolu_moments}
\begin{aligned}
\partial_t \langle M_\bullet^p\rangle & = \partial_t \,\int_0^\infty{\rm d}M_\bullet\, N(M_\bullet,t)\, M_\bullet^p =\\
&\\
& = \int_0^\infty{\rm d}M_\bullet'\, K(M_\bullet,M_\bullet',t)\, N(M_\bullet,t)\, N(M_\bullet',t)\,[(M_\bullet+M_\bullet')^p-M_\bullet^p-M_\bullet'^p]\;,
\end{aligned}
\end{equation}
where $p$ is an integer (e.g., \cite{Leyvraz2003,Smoluchowski1916,Cavaliere1992}). The above equation makes evident that $\partial_t \langle M_\bullet\rangle = 0$ and $\partial_t \langle M_\bullet^2\rangle>0$; in other words, the coagulation processes does not cause a drift in the mass distribution, but just enhances its dispersion by redistributing masses from the low to the high mass end. We mention that there is an exception to the mass conservation $\langle\dot M_\bullet\rangle = 0$ found above: when the kernel dependence on the mass is sufficiently strong, like in the well-known case of a multiplicative kernel $K(M_\bullet,M_\bullet')\propto M_\bullet\, M_\bullet'$, then a sort of phase transition occurs, in that some mass may be lost in a finite time due to the appearance of clusters with infinite mass \cite{Cavaliere1991}. Such a phenomenon is called `gelation' and has attracted a lot of interest in the scientific community for its practical applications to the physics of aereosol, polymers and molecular dynamics; however, this is not of our concern here (for a review, see \cite{Leyvraz2003}).

Note that the quantity $K(M_{\bullet,1},M_{\bullet,2})\, N(M_{\bullet,1})\,N(M_{\bullet,2})\equiv {\rm d}^2\mathcal{R}/{\rm d}M_{\bullet,1}\,{\rm d}M_{\bullet,2}$ just represents a merging rate density per bins of the two merging masses $M_{\bullet,1}$ and $M_{\bullet,2}$. However, in the present context it is more convenient to work with the merger rate density ${\rm d}^2\mathcal{R}_{\bullet\bullet\rightarrow \bullet}/{\rm d}M_{\bullet,1+2}\,{\rm d}q$ per bin of total (descendant) mass $M_{\bullet,1+2}\equiv M_{\bullet,1}+M_{\bullet,2}$ and mass ratio $q\equiv \min(M_{\rm \bullet,1};M_{\bullet,2})$ $/\max(M_{\rm \bullet,1};M_{\bullet,2})$, i.e. the ratio between the lesser and the more massive merging BHs (by definition). This can be achieved by a suitable change of variables in each term of the Smoluchowski equation, taking into account that the the Jacobian of the transformation is just $M_{\bullet,1+2}/(1+q)^2$.
Identifying the two appropriate merging masses in each integral of Equation (\ref{eq|Smolu}), and splitting the destruction term into two parts to always have mass ratios $q<1$, the overall Smoluchowski equation (\ref{eq|Smolu}) can be rewritten in a rather compact form as (see also \cite{SatoPolito2025})
\begin{equation}\label{eq|Smolu_newvar}
\begin{aligned}
[\partial_t N(M_\bullet,t)]_{\bullet\bullet\rightarrow \bullet} = \int_0^1{\rm d}q\, \frac{{\rm d}^2 \mathcal{R}_{\bullet\bullet\rightarrow \bullet}}{{\rm d}M_\bullet\,{\rm d}q}_{\big|M_\bullet} & - \int_0^1{\rm d}q\, (1+q)\, \left[ \frac{{\rm d}^2 \mathcal{R}_{\bullet\bullet\rightarrow \bullet}}{{\rm d}M_\bullet\,{\rm d}q}_{\big|M_\bullet\,(1+q)}+ \right.\\
&\\
&\left.+ q\, \frac{{\rm d}^2 \mathcal{R}_{\bullet\bullet\rightarrow \bullet}}{{\rm d}M_\bullet\,{\rm d}q}_{\big|M_\bullet\,(1+q)/q}\right]\; .
\end{aligned}
\end{equation}

Simulations indicate that the dependencies on mass, mass ratio and redshift of the merger rate densities per descendant dark matter halo or per descendant galaxy can be factorized \cite{Fakhouri2009,Rodriguez2016}, so we make the simplifying assumption that the same holds for the hosted supermassive BHs. Therefore we parameterize the merger rate density per BH as
\begin{equation}\label{eq|mergratebase}
\frac{1}{N(M_\bullet,t)}\,\frac{{\rm d}^2\mathcal{R}_{\bullet\bullet\rightarrow \bullet}}{{\rm d}M_\bullet\,{\rm d}q}(M_\bullet,q,t) \simeq
\mathcal{R}_{\bullet\bullet\rightarrow \bullet}(t)\, \left(\frac{M_\bullet}{10^{10}\, M_\odot}\right)^\eta\, p(q)\; ;
\end{equation}
where $\eta$ is a parameter specifying the mass dependence, ${\rm d}p/{\rm d}q$
is the probability density distribution of mass ratios, and $\mathcal{R_{\bullet\bullet\rightarrow \bullet}}(t)$ is a redshift dependent normalization. Then the coagulation term turns into:
\begin{equation}\label{eq|Smolu_final}
\begin{aligned}
[\partial_t N(M_\bullet,t)]_{\bullet\bullet\rightarrow \bullet} = \mathcal{R}_{\bullet\bullet\rightarrow \bullet}(t)\, \left(\frac{M_\bullet}{10^{10}\, M_\odot}\right)^\eta\, & \Bigg\{N(M_\bullet,t)  - \int_0^1{\rm d}q\;p(q)\,(1+q)^{1+\eta}\,\left[N(M_\bullet,t)_{\big|M_\bullet\,(1+q)}\right.\\
&\\
&\left.+q^{1-\eta}\, N(M_\bullet,t)_{\big|M_\bullet\,(1+q)/q}\right]\Bigg\}\;.
\end{aligned}
\end{equation}
We parameterize the redshift-dependent merger rate normalization in terms of a rescaling of the galaxy merger rate as estimated from galaxy pair counts observed by \texttt{JWST} (see \cite{Casteels2014,Duan2025,Puskas2025}). Specifically, we take
\begin{equation}\label{eq|mergrate}
\mathcal{R}_{\bullet\bullet\rightarrow \bullet}(t) = f_{\bullet\bullet\rightarrow \bullet}\, \, \mathcal{R}_0\,(1+z)^\omega\, e^{-\tau\,(1+z)} 
\end{equation}
in terms of a scaling factor $f_{\bullet\bullet\rightarrow \bullet}$ and of the parameters $\mathcal{R}_0\approx 0.013$ Gyr$^{-1}$, $\omega\approx 3.4$ and $\tau\approx 0.14$ set by \cite{Duan2025}. This choice is appropriate since it would be impossible to constrain the detailed redshift dependence of the BH merger rate relying only on the data concerning BH demographics and clustering, and on the mild integrated redshift dependence of the stochastic background in GWs (see below). 

We anticipate that in our analysis $f_{\bullet\bullet\rightarrow\bullet}$ is not restricted to be less than unity, although it would be somewhat nonphysical to obtain a BH merger rate larger than the galactic one. However, this is done to account for observational scatter or possible systematics offsets in the galaxy merger rate. In fact, the latter is obtained combining the observed galaxy pairs fraction with a merger timescale extracted from numerical simulations, and thus inevitably subject to some uncertainties. Note that a different normalization of the merger rate will lead to a recalibration of $f_{\bullet\bullet\rightarrow \bullet}$ but will marginally affect the conclusions of the present work.

In addition, note that in the above we are neglecting any possible time delay between the merger of the BHs and that of the host galaxies, which constitutes the standard assumption in the analysis of the GW background \cite{Sesana2016,Agazie2023b}; to wit, this maximizes the contribution of binary mergers. 

Finally, following \cite{Kelley2017,SatoPolito2024} we take the mass ratio distribution $p(q)\propto q^{-\delta}\, \theta_{\rm H}[q\geq q_{\rm min}]$ as a power-law with index $\delta$ extending down to a minimum mass ratio $q_{\rm min}$, held fixed to the reference values $\delta = -1$ and $q_{\rm min}=0.1$. In fact, the dependence of the results on the mass ratio distribution is very weak, so there is no point in trying to fit the related parameters.

\subsection{BH demographics}\label{sec|demographics}

The direct output of the continuity equation is constituted by the BH demographics, i.e. the BH mass function at any cosmic time. This is found by solving the continuity equation:
\begin{equation}\label{eq|continuity}
\begin{aligned}
\partial_t\, N(\log M_\bullet,t) &= -\frac{\partial_{\ln M_\bullet}}{t_{\rm Edd}}\,\Bigg[\int{\rm d}\log \lambda\; p(\log\lambda)\, \lambda\, \frac{1-\epsilon-\epsilon_{\rm kin}}{\epsilon}\,N_{\rm AGN}(\log M_\bullet,t)\Bigg]+\\
&\\
&+\mathcal{R}_{\bullet\bullet\rightarrow \bullet}(t)\, \left(\frac{M_\bullet}{10^{10}\,M_\odot}\right)^\eta\, \Bigg\{N(\log M_\bullet,t)+ \\
&\\
& - \int_0^1{\rm d}q\;p(q)\,(1+q)^\eta\,\left[N(\log M_\bullet,t)_{|M_\bullet(1+q)}+q^{2-\eta}\,N(\log M_\bullet,t)_{|M_\bullet(1+q)/q}\right]\Bigg\}~.
\end{aligned}
\end{equation}
Note that although this is formally a partial integro-differential equation, the integrals only involve the mass function evaluated at the running time, so as to allow the exploitation of standard numerical techniques. In more detail, we solve Equation (\ref{eq|continuity}) forward in cosmic time on discrete grids in redshift $z\in [0,z_{\rm ini}]$ and BH masses $\log M_{\bullet}\, [M_\odot]\in [5,11]$ with an explicit Runge-Kutta method of eight order. 

Fiducially, we initialize the computation at $z_{\rm ini}\sim 7$ which is the maximal redshift at which measurements of the AGN luminosity function by \cite{Shen2020} do not require extrapolation, and there we assume as initial condition a mean duty cycle $\langle\delta\rangle(M_\bullet,z_{\rm ini}) = 0.5$ independent of mass \cite{Shankar2013}. Then by definition of the duty cycle one has $N_{\rm AGN}(\log M_\bullet,z_{\rm ini}) = N(\log M_\bullet,z_{\rm ini})\,\langle\delta\rangle = N(\log M_\bullet,z_{\rm ini})/2$ and hence, given the observed AGN luminosity function $N(L_{\rm AGN},z_{\rm ini})$, one can determine the initial BH mass function $N(\log M_\bullet,z_{\rm ini})$ just by solving Equation (\ref{eq|conv}).
To explore the dependence of our results on the initial condition, we try to use $z_{\rm ini}\sim 10$ by extrapolating the observed AGN luminosity functions. We also try to assume a different initial duty cycle in the range $\langle\delta\rangle(M_\bullet,z_{\rm ini}) = 0.1-1$. Finally, we try a narrowly peaked mass function or the mass function stemming from various light-seed population models \cite{Volonteri2010,Sicilia2022a}. Basically, we have found that as for the determination of the BH mass function at $z\lesssim 4$ and below, the initial redshift and reasonable initial conditions  are marginally relevant since the accumulated mass from accretion soon largely exceeds that in the original BH population (see also \cite{Aversa2015,Sicilia2022,Roberts2025}), and the impact of BH mergers on the latter at high redshift will turn out to be relatively minor.

\subsection{BH Clustering}\label{sec|clustering}

A further observable to better constrain the parameters of our semi-empirical framework, and in particular to contribute in reducing the degeneracy between the parameters is to exploit the large-scale BH clustering \cite{Shankar2013,Shankar2020}. Specifically, we will be interested in comparing our model outcomes to data concerning the bias $b(M_\bullet,z)$, as provided at relatively low redshifts $z\lesssim 0.25$ by optical and X-ray observations of AGNs with robust BH mass determinations \cite{Krumpe2015,Krumpe2023}. 

In the context of our semi-empirical framework, the BH clustering signal can be computed by combining the BH mass function output from the continuity equation to the mass function and bias of dark matter halos as extracted from $N-$body simulations. In more detail, the procedure is the following. First of all, we employ the standard abundance matching technique \cite{Shankar2004,Moster2013} to relate BH and halo masses in a nearly model-independent way, the only assumption being the existence of a monotonic mapping $M_{\rm H}(M_\bullet,z)$ between the two variables, at least on statistical grounds. In particular, we use the version of abundance matching by \cite{Aversa2015}
that includes the effect of the scatter in the halo-to-BH mass relation. This boils down to require the cumulative number densities in BHs and halos to match as:
\begin{equation}\label{eq|abma}
\int^{+\infty}_{\log M_\bullet}{\rm d}\log M_\bullet'\;N(\log M_\bullet',z) = \int_{-\infty}^{+\infty}{\rm d}\log M_{\rm H}'\; N(\log M_{\rm H}',z)\times \frac{1}{2}\, {\rm erfc}\Bigg[\frac{\log M_{\rm H}-\log M_{\rm H}'}{\sqrt{2}\sigma_{\log M_{\rm H}}}\Bigg]\;,
\end{equation}
and solve for $M_{\rm H}(M_\bullet,z)$; here the scatter of the relation $\sigma_{\log M_{\rm H}}$ dex is a fitting parameter. The integrand on the left hand side encases the BH mass function, that we obtain by solving the continuity equation. The right hand side involves the $\Lambda$CDM halo mass function $N(\log M_{\rm H},z)$, that we take from the $N-$body simulations by \cite{Tinker2008}. Note that this procedure is also supported by the recent pairwise residuals analysis by \cite{Shankar2025}, which has highlighted that the correlation between BH and host halo mass appears as strong as the one with velocity dispersion, as corroborated by direct fitting (e.g., \cite{Ferrarese2002}) and clustering analysis (e.g., \cite{Powell2022}). Finally, we employ the large-scale bias $b(M_{\rm H},z)$ of dark matter halos from the $N-$body simulations by \cite{Tinker2010}, and convolve it with the halo mass vs. BH mass relation (properly taking into account its dispersion $\sigma_{\log M_{\rm H}}$) to obtain the BH bias $b(M_\bullet,z)$ that can be then directly compared with clustering data. To speed up the above computation we actually exploit the accurate fitting formulas for the halo mass function\footnote{We also include the correction for sub-halos in the halo mass function. However, this becomes relevant only at halo masses $M_{\rm H}\lesssim 10^{12}\, M_\odot$ and thus it is found to play a minor role in the present context.} and for the halo bias from the aforementioned $N-$body simulations, as implemented in the \texttt{COLOSSUS} package \cite{Diemer2018}.

It is worth stressing that the procedure described 
above does not betray the data-driven spirit of our semi-empirical framework, because it does not require any major theoretical assumption apart from the underlying cosmological model (standard $\Lambda$CDM in our case) and the reasonable hypothesis that on statistical average larger BHs are associated to more massive halos. Note also the different  way through which clustering is obtained within ab-initio models. These are also based on the dark matter halo catalogs (containing the halo clustering information) extracted from $N-$body simulations, but then the mapping between BH and halo masses requires to make substantial theoretical assumptions and to model the very complex baryonic physics involved in the formation and evolution of BHs. Such a challenging task is instead completely bypassed by our semi-empirical, data-driven framework based on the continuity equation. 

\subsection{Stochastic GW background}\label{sec|background}

Binary mergers of supermassive BH throughout cosmic history can produce a stochastic backround of GW with typical nano-Hz frequency, that is within the reach of present PTA experiments. In the context of our semi-empirical framework, this can be easily computed from the outputs of the continuity equation, and in particular from the BH mass function and the BH merger rate. Specifically, the characteristic strain $h_c$ of the stochastic GW background from binary BH mergers (in circular orbits) as a function of the observed frequency $f$ is given by \cite{Phinney2001}
\begin{equation}\label{eq|GWstrain}
\begin{aligned}
h_c^2(f) & = \frac{4\, G^{5/3}}{3\,\pi^{1/3}\, c^2\, f^{4/3}}\, \int\frac{{\rm d}t}{(1+z)^{1/3}}\, \int{\rm d}M_\bullet\, M_\bullet^{5/3}\int{\rm d}q\, \frac{q}{(1+q)^2}\,\frac{{\rm d}^2\mathcal{R}_{\bullet\bullet\rightarrow \bullet}}{{\rm d}M_\bullet\,{\rm d}q} = \\
&\\
& = \frac{4\, G^{5/3}}{3\,\pi^{1/3}\, c^2\, f^{4/3}}\, \int{\rm d}t\, \frac{\mathcal{R}_{\bullet\bullet\rightarrow \bullet}(t)}{(1+z)^{1/3}}\, \int{\rm d}M_\bullet\, M_\bullet^{5/3}\, \left(\frac{M_\bullet}{10^{10}\, M_\odot}\right)^\eta\,N(M_\bullet,t)\int{\rm d}q\; p(q)\, \frac{q}{(1+q)^2}\;.
\end{aligned}
\end{equation}
where $N(M_\bullet,t)$ is the BH mass function obtained from solving the continuity Equation (\ref{eq|continuity}) and $\mathcal{R}_{\bullet\bullet\rightarrow\bullet}$ is given by Equation (\ref{eq|mergrate}). Note that very often results of nano-Hz GW experiments are quoted in terms of a normalization $\mathcal{A}$ for the characteristic strain \cite{Jenet2006} written in the form $h_c(f) = \mathcal{A}\, (f/{\rm yr}^{-1})^{-2/3}$, where the constant $\mathcal{A}$ can be computed easily from the expression above. The $f^{-2/3}$ scaling holds for circular orbits, and it can actually be altered by eccentricity or by environmental effects (in that they tend to somewhat reduce the strain and flatten its frequency dependence; see \cite{Fastidio2025}), but these go beyond the scope of the present paper. 

\section{Data and analysis}\label{sec|data}

We now turn to constrain the parameters of our semi-empirical framework by comparing its outcomes with various datasets. Specifically, we exploit local determinations of the BH mass function, of the BH clustering, and a recent estimate of the nano-Hz GW background. 

\begin{figure}[t!]
    \centering
    \includegraphics[width=1.\textwidth]{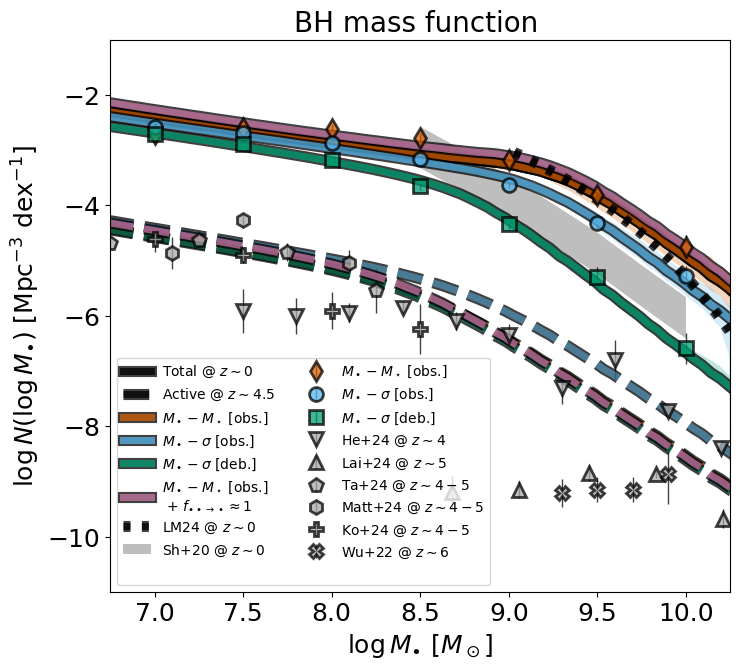}
    \caption{The supermassive BH mass function. Colored symbols refer to the local BH mass function determined from SMF and observed $M_\bullet-M_\star$ relation (green squares), from VDF and observed $M_\bullet-\sigma$ relation (blue circles), and from VDF and debiased $M_\bullet-\sigma$ relation (red rhomboids). These determinations bracket the overall systematics uncertainty on the estimates of the local BH mass function. For comparison, the dotted line is the recent determination based on the $M_\bullet-M_\star$ scaling relation by \cite{Liepold2024}; the gray shaded area is the determination by \cite{Shen2020}, obtained convolving the AGN luminosity function with the AGN luminosity vs. Eddington ratio relationship from SDSS \cite{Shen2009,Nobuta2012} and further correcting for obscured AGNs and inactive BHs. Grey symbols display recent estimates of the active BH mass function at $z\sim 4-6$ including BAL quasars and little red dots by \cite{He2024} (reverse triangles), \cite{Lai2024} (triangles), \cite{Taylor2025} (pentagons), \cite{Matthee2024} (hexagons), \cite{Kokorev2024} (plus signs), and \cite{Wu2022} (crosses). Solid lines with shaded areas illustrate the outcomes (median and $2\sigma$ credible intervals) for the local BH mass function reconstructed from our semi-empirical framework and MCMC analysis, with the same color-code as in previous Figure, while dashed lines illustrate the corresponding active BH mass functions at $z\sim 4.5$.}
    \label{fig|BHMF}
\end{figure}

As mentioned in Section \ref{sec|intro}, the local BH mass function cannot be estimated directly, since the number of robustly measured BH masses is small (less than a hundred). Thus this is routinely obtained via the convolution of a galaxy statistics (measured on a much large sample of galaxies) with a scaling relationship between BH mass and galaxy properties (calibrated on the small sample of directly measured BH masses). Unfortunately, this procedure brings in a certain degree of uncertainty in the determination of the local BH mass function, which is important to take into account. To this purpose, we consider three cases. In the first case, the local BH mass function is obtained by combining the local stellar mass function or SMF $N(M_\star)$ by \cite{Moffett2016} with the $M_\bullet-M_\star$ relationship for early-type galaxies by \cite{Savorgnan2016,Shankar2020}. In the second case, the local BH mass function is obtained by combining the local galaxy velocity dispersion function or VDF $N(\sigma)$ by \cite{Sheth2003} and the observed $M_\bullet-\sigma$ relationship by \cite{McConnell2013}. Finally, in the third case the local BH mass function is obtained by combining the VDF with the 
`debiased' version of the $M_\bullet-\sigma$ by \cite{Shankar2016}; this is meant to constitute a solid lower limit to the local demography, as pointed out by the recent results in \cite{Shankar2025}.
Such binned local BH mass functions are reported in Table \ref{tab|data}. Our analysis will be conducted independently for each of these three cases, which confidently bracket the overall systematic uncertainty\footnote{Note that using the SMF and the intrinsic $M_\bullet-M_\star$ relationship by \cite{Shankar2016} produces a local BH mass function consistent with the combination VDF + intrinsic $M_\bullet-\sigma$, as shown explicitly by \cite{Shankar2020}.}. They also span the range covered by other literature estimates from different galaxy statistics and/or relationships. For example, the recent BH mass function by \cite{Liepold2024} strikes an intermediate course between our cases based on the observed $M_\bullet-\sigma$ and $M_\bullet-M_\star$ relations (see dotted line in Figure \ref{fig|BHMF}). As another example, the determination by \cite{Shen2020}, obtained convolving the AGN luminosity function with the AGN luminosity vs. Eddington ratio relationship from SDSS \cite{Shen2009,Nobuta2012} and further correcting for obscured AGNs and inactive BHs, is in between our cases based on the observed and the debiased $M_\bullet-\sigma$ relations (see shaded area in Figure \ref{fig|BHMF}).

As to BH clustering, we rely on the measurements of the large-scale bias $b(M_\bullet)$ as a function of BH mass at $z\sim 0.25$ by \cite{Krumpe2015,Krumpe2023} and at $z\sim 0.04$ by \cite{Powell2018,Allevato2021}. Note that a wealth of additional datasets (e.g., \cite{Shen2009,Allevato2011, White2012,Shen2013,Allevato2014,Allevato2016,Mountrichas2016,Magliocchetti2017,Hale2018,Powell2018,Eftekharzadeh2019,Ross2020,Krishnan2020,Mazumder2022,Hale2025}) are available at different redshifts for the clustering of AGNs as a function of luminosity from optical, X-ray and radio selected sample. However, we cannot rely on these datasets to independently test our framework. This is because we used as an input the AGN luminosity function, hence the relation between AGN luminosity and halo mass (obtained via abundance matching, see Section \ref{sec|clustering}) and the ensuing clustering signal as a function of AGN luminosity is reproduced by construction. Focusing instead on the clustering signal as a function of BH mass can independently constrain the halo to BH mass relationship and hence ultimately the BH mass function, which is the main output of our framework (see also \cite{Shankar2020,Allevato2021}).

As to the GW background, we rely on the most recent estimate from PTA by the \texttt{NANOGrav} collaboration \cite{Agazie2023}, in terms of the strain amplitude parameter $\mathcal{A} = (2.4\pm 0.7)\times 10^{-15}$ expressed by Equation (\ref{eq|GWstrain}). We stress again that an added value of the present work is in attempting to fit \emph{jointly} all these observables. In Appendix \ref{app|tests} we will also test our model outcomes when excluding the PTA measurement from the likelihood, i.e. fitting only for BH demographics and clustering and then compare the resulting GW background with data. 

We will also confront the outcomes of our semi-empirical framework with other auxiliary datasets not included in the fitting procedure: measurements of the local duty cycle  $\langle\delta\rangle(M_\bullet)$ of AGN activity as a function of BH mass (taking into account appropriate luminosity threshold) by 
\cite{Kauffmann2003,Goulding2010,Schulze2010}; estimates of the active BH mass function at high-redshift $z\sim 4-6$, including BAL quasars and little red dots by \cite{Wu2022,Lai2024,He2024,Matthee2024,Taylor2025,Kokorev2024,Geris2025}; BH vs. halo mass relationship inferred from different methods \cite{Marasco2021,Krumpe2015,Robinson2021,Li2024}; evolution with redshift of the average Eddington ratio for various kinds of AGNs \cite{Vestergaard2009,Nobuta2012,Dai2014,Farina2022,He2024,Lai2024,Fan2023,Harikane2023,Greene2024,Maiolino2024,Matthee2024}. Being affected by appreciable uncertainties and possibly by systematics or observational biases, these datasets are not employed to fit the parameters of our semi-empirical framework, but rather are exploited as \emph{a-posteriori} validation.
 
{\renewcommand{\arraystretch}{1.15}\setlength{\tabcolsep}{2.5pt}
\begin{table}[htbp]
\footnotesize
\centering
\begin{tabular}{ccccccccccccc}
\hline
\hline
$\log M_\bullet$ [$M_\odot$] && \multicolumn{7}{c}{$\log N(\log M_\bullet)$ [Mpc$^{-1}$ dex$^{-1}$]} \\
\hline
&& $M_\bullet-M_\star$ [obs.] &&& $M_\bullet-\sigma$ [obs.]  &&& $M_\bullet-\sigma$ [deb.]\\
$7.00$ && $-2.71\pm 0.06$ &&& $-2.59\pm 0.04$ &&& $-2.70\pm 0.05$\\
$7.50$ && $-2.60\pm 0.06$ &&& $-2.69\pm 0.04$ &&& $-2.88\pm 0.05$\\
$8.00$ && $-2.62\pm 0.08$ &&& $-2.87\pm 0.05$ &&& $-3.18\pm 0.05$\\ 
$8.50$ && $-2.79\pm 0.10$ &&& $-3.17\pm 0.06$ &&& $-3.65\pm 0.07$\\
$9.00$ && $-3.18\pm 0.12$ &&& $-3.63\pm 0.07$ &&& $-4.34\pm 0.11$\\
$9.50$ && $-3.83\pm 0.16$ &&& $-4.32\pm 0.11$ &&& $-5.30\pm 0.18$\\
$10.0$ && $-4.76\pm 0.23$ &&& $-5.28\pm 0.18$ &&& $-6.59\pm 0.28$\\
\hline
\hline
\end{tabular}
\caption{The local BH mass functions exploited in our analysis. These include the determination from the stellar mass function (SMF) from \cite{Moffett2016} plus the $M_\bullet-M_\star$ relationship from \cite{Savorgnan2016,Shankar2020} , the velocity dispersion function (VDF) from \cite{Sheth2003} plus the $M_\bullet-\sigma$ relationship in its observed [obs.] version from \cite{McConnell2013} or in its debiased [deb.] version from \cite{Shankar2016}. See text for more details.}
\label{tab|data} 
\end{table}} 

For the analysis we adopt a Bayesian framework, characterized by the parameter set $\theta\equiv \{\epsilon_{\rm thin},\log\lambda_0,\xi,\eta,\log f_{\bullet\bullet\rightarrow\bullet},\sigma_{\log M_{\rm H}}\}$. We implement a Gaussian log-likelihood
\begin{equation}\label{eq|likelihood}
\ln \mathcal{L}(\theta) = -\chi^2(\theta)/2~,
\end{equation}
where the chi-square $\chi^2(\theta)=\sum_{ij}\,[\mathcal{M}(M_{\bullet,i},z_i|\theta)-\mathcal{D}(M_{\bullet,i},z_i)]\,\mathcal{C}_{ij}^{-1}\, [\mathcal{M}(M_{\bullet,j},z_j|\theta)-\mathcal{D}(M_j,z_j)]$ is obtained by comparing our model expectations $\mathcal{M}(M_{\bullet,i},z_i|\theta)$ to the data values $\mathcal{D}(M_{\bullet,i},z_i)$, summing over different observables at their respective BH masses $M_{\bullet,i}$ and redshifts $z_i$ and taking into account the variance-covariance matrix $\mathcal{C}_{ij}$ among different bins (here assumed diagonal). We adopt flat priors $\pi(\theta)$ on the parameters $\epsilon_{\rm thin}\in (0.06,0.42)$, $\log\lambda_0\in [-3,1]$, $\xi\in [0,3]$, $\eta\in [-1,1]$, $\sigma_{\log\lambda}\in (0,1]$, $\log f_{\bullet\bullet\rightarrow\bullet}\in [-3,1]$, $\sigma_{\log M_{\rm H}}\in (0,1]$. Note that $f_{\bullet\bullet\rightarrow\bullet}$ is not hard bounded by one, to take into account possible uncertainties in the galaxy merger rate. We will also consider a run where the supermassive BH merger rate follows unbiasedly the galaxy merger rate by setting a narrow gaussian prior on $\log f_{\bullet\bullet\rightarrow\bullet}$ centered in zero and with dispersion $0.1$ dex. The overall set of parameters, the related equation in the main text where these first appear, and the adopted priors are reported in Table \ref{tab|param}.

We sample the parameter posterior distributions $\mathcal{P}(\theta) \propto \mathcal{L}(\theta)\,\pi(\theta)$ via the MCMC Python package \texttt{emcee} \cite{emcee}, running it with $10^5$ steps and $100$ walkers; each walker is initialized with a random position extracted from the priors discussed above. 
To speed up the computation, the active BH mass function is pre-determined solving Equation (\ref{eq|convo}) on a grid of redshift and Eddington ratio, and then its parameters are interpolated on the fly during the MCMC run. 
To more efficiently sample the parameter space, we adopt a mixture of differential evolution and snooker moves of the walkers, in proportion of $0.8$ and $0.2$ respectively, that emulates a parallel tempering algorithm. 
After checking the auto-correlation time, we remove the first $50\%$ of the flattened chain to ensure burn-in.

{\renewcommand{\arraystretch}{1.5}  \setlength{\tabcolsep}{2.5pt}
\begin{table}[htbp]
\footnotesize
\centering
\begin{tabular}{lcccccc}
\hline
\hline
Parameter & Prior & Equation\\
\hline
$\log\lambda_0$ & $\mathcal{U}[-3,1]$ & \ref{eq|lambda}\\
$\xi$ & $\mathcal{U}[0,3]$ & \ref{eq|lambda}\\
$\epsilon_{\rm thin}$ & $\mathcal{U}(0.06,0.42)$ & \ref{eq|epsilon}\\
$\eta$ & $\mathcal{U}[-1,1]$ & \ref{eq|mergratebase}\\
$\log f_{\bullet\bullet\rightarrow\bullet}$ & $\mathcal{U}[-3,1]$ or $\mathcal{N}(0,0.1)$& \ref{eq|mergrate}\\
$\sigma_{\log M_{\rm H}}$ & $\mathcal{U}(0,1]$ & \ref{eq|abma}\\
\hline
\hline
\end{tabular}
\caption{Parameter set exploited in our Bayesian analysis. The first column lists the parameter symbol, the second column shows the related prior (the symbol $\mathcal{U}$ stands for a uniform distribution between the arguments, and $\mathcal{N}$ for a normal distribution with mean and dispersion given by the arguments), and the third column recall the Equation of the main text where the parameter appears.}
\label{tab|param} 
\end{table}} 

{\renewcommand{\arraystretch}{2.}  \setlength{\tabcolsep}{2.5pt}
\begin{table}[htbp]
\footnotesize
\centering
\begin{tabular}{lccccccccccccc}
\hline
\hline
Parameter & & $M_\bullet-M_\star$ [obs.] & & $M_\bullet-\sigma$ [obs.] & & $M_\bullet-\sigma$ [deb.] & & $M_\bullet-M_\star$ [obs.] $+f_{\bullet\bullet\rightarrow \bullet}\approx 1$ \\
\hline
$\epsilon_{\rm thin}$ && $0.066^{+0.003}_{-0.008}$ && $0.11^{+0.01}_{-0.01}$ && $0.34^{+0.02}_{-0.03}$ && $0.066^{+0.003}_{-0.008}$\\
$\log\lambda_0$ && $-2.25^{+0.26}_{-0.23}$ && $-1.43^{+0.22}_{-0.15}$ &&  $-0.62^{+0.12}_{-0.08}$ &&  $-2.22^{+0.27}_{-0.22}$\\
$\xi$ && $2.57^{+0.64}_{-0.64}$ && $1.03^{+0.38}_{-0.61}$ && $0.36^{+0.06}_{-0.42}$ && $2.53^{+0.58}_{-0.68}$\\
$\eta$ && $0.59^{+0.44}_{-0.68}$ && $0.64^{+0.48}_{-0.66}$ && $0.65^{+0.58}_{-0.83}$ && $0.52^{+0.33}_{-0.38}$\\
$\log f_{\bullet\bullet\rightarrow\bullet}$ && $-0.77^{+0.8}_{-0.5}$ && $-0.85^{+0.9}_{-0.6}$ && $-1.4^{+1.1}_{-0.9}$ && $-0.04^{+0.09}_{-0.09}$\\
$\sigma_{\log M_{\rm H}}$ && $0.68^{+0.04}_{-0.03}$ && $0.58^{+0.04}_{-0.04}$ && $0.25^{+0.08}_{-0.07}$ && $0.68^{+0.04}_{-0.04}$\\
\hline
$\chi^2_r$ && $1.22$ && $1.24$ && $1.25$ && $1.47$\\
\hline
\hline
\end{tabular}
\caption{Results of our Bayesian analysis in terms of marginalized posterior estimates. For every parameter (listed in the first column) of our semi-empirical framework we report the median and the $2\sigma$ credible interval. The different columns refer to the different determinations of the local BH mass functions discussed in the main text. The last line lists the reduced $\chi^2_r$ of the overall fit in the various cases.}
\label{tab|results} 
\end{table}}

\subsection{Results and discussion}\label{sec|results}

\begin{figure}[t!]
    \centering
    \includegraphics[width=1.\textwidth]{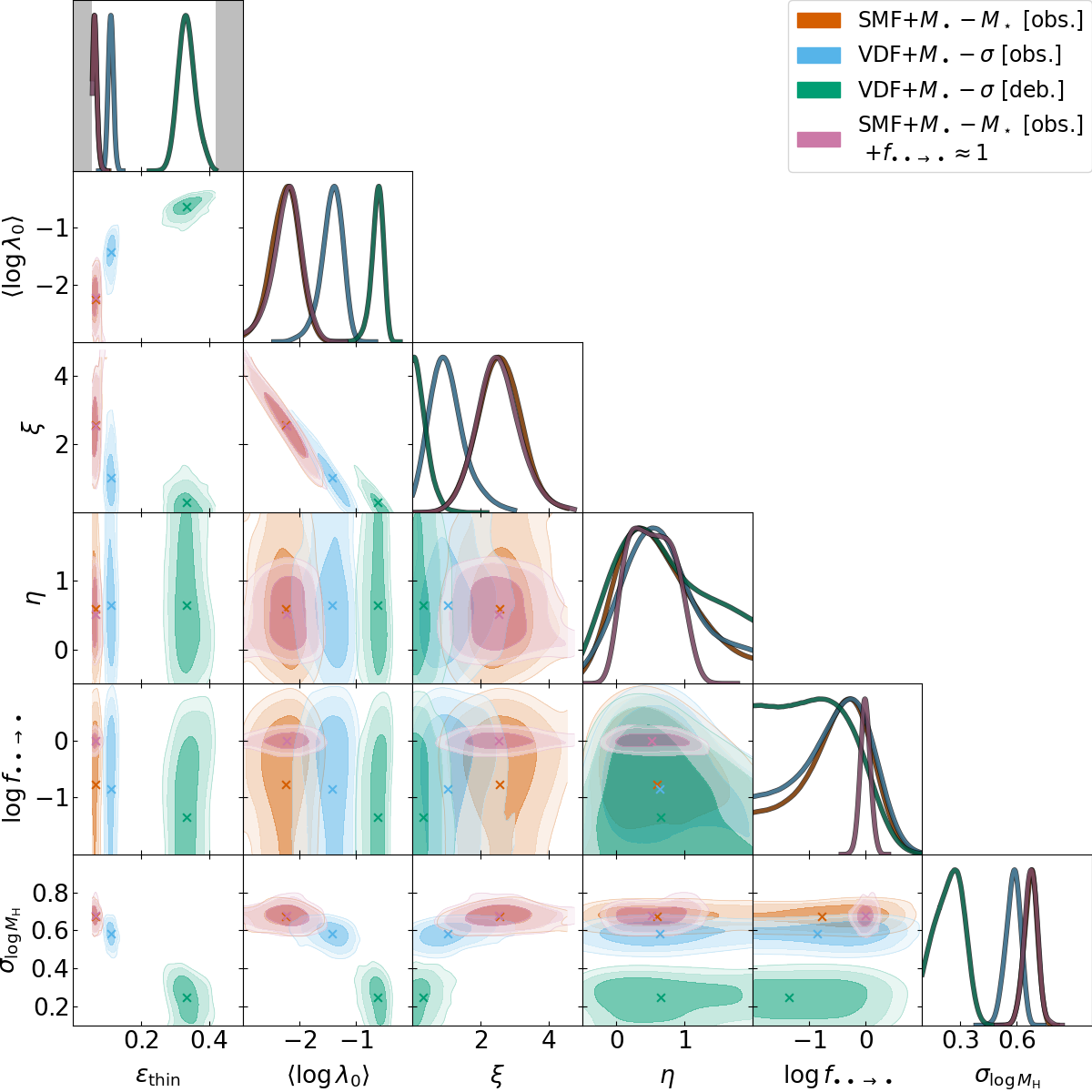}
    \caption{MCMC posterior distributions for the parameters exploited in our semi-empirical framework: radiative efficiency of the thin disk phase $\epsilon_{\rm thin}$; mean Eddington ratio $\langle\log \lambda_0 \rangle$ and its redshift evolution parameter $\xi$; mass-dependence parameter $\eta$ in the supermassive BH merger rate and  rescaling parameter $f_{\bullet\bullet\rightarrow \bullet}$ of the BH to galaxy merger rate; scatter $\sigma_{\log M_{\rm H}}$ in the abundance matching relationship between halo and BH mass. Coloured contours/lines refer to the analysis with different determinations of the local BH mass function: orange is for that obtained combining the local SMF with the $M_\bullet-M_\star$ relation, and magenta adds a narrow prior around $f_{\bullet\bullet\rightarrow\bullet}\approx 1$; blue is for that obtained combining the VDF with the observed $M_\bullet-\sigma$ relation; green is for that obtained combining the VDF with the debiased $M_\bullet-\sigma$ relation. The marginalized distributions are in arbitrary units (normalized to unity at their maximum value). Colored crosses mark bestfit values. In the uppermost panel, the gray shaded areas highlight non-physical values of the thin-disk radiative efficiency $\epsilon_{\rm thin}$ which are excluded by hard priors.}
    \label{fig|MCMC}
\end{figure}

The results of our Bayesian analysis are displayed in Figure \ref{fig|MCMC} and are reported in Table \ref{tab|results}. Specifically, in Figure \ref{fig|MCMC} we illustrate the MCMC posterior distributions on 
our parameter set, for the different determinations of the local BH mass function (color-coded); the crosses highlight the positions of the bestfit and the marginalized distributions are normalized to unity at their maximum. In Table \ref{tab|results} we summarize the marginalized posterior estimates of the parameters, in terms of their median values and $1\sigma$ credible intervals.

\subsubsection{Parameter estimates}

The radiative efficiency $\epsilon_{\rm thin}$ is a well determined parameter, though it is very sensitive to the adopted local BH mass function determination: values $\epsilon_{\rm thin}\approx 0.06$ close to the lower bound allowed for a thin disk are required when adopting the BH mass function based on the observed $M_\bullet-M_\star$ relationship; values $\epsilon_{\rm thin}\approx 0.1$ around the usually adopted one for the average population of AGNs (e.g., \cite{Davis2011,Trakhtenbrot2017}) are found when basing on the observed $M_\bullet-\sigma$ relationship; extreme values $\epsilon_{\rm thin}\approx 0.3$ close to the upper bound allowed for a thin disk are required when adopting the BH mass function based on the debiased $M_\bullet-\sigma$ relationship. These variations are easily understood basing on the classic Soltan \cite{Soltan1982,Salucci1999} argument. As it is more easily seen in  the simple case of Equation (\ref{eq|easycont}), the local BH mass density $\rho_{\bullet}\equiv \int{\rm d}\log M_\bullet\, M_\bullet\, N(\log M_\bullet,t)$ is essentially determined by the integral over cosmic time of the AGN luminosity density $\rho_\bullet \propto \frac{1-\epsilon-\epsilon_{\rm kin}}{\epsilon}\, \int{\rm d}t\, \int{\rm d}\log L_{\rm AGN}\, L_{\rm AGN}\, N(\log L_{\rm AGN},t)$ scaled by the inverse of the radiative efficiency $\epsilon^{-1}$. Thus BH mass function determinations featuring a higher normalization at the knee, which implies a larger BH mass density $\rho_\bullet$, require lower average efficiencies. 

The values of the average efficiency estimated here could have far reaching physical consequences, since in the standard picture of disk accretion these can be traced back to the spinning state of the supermassive BHs \cite{Bardeen1972,Zhang2019}. For example, if the local BH mass function from the debiased $M_\bullet-\sigma$ relationship would be the true one, the related high efficiencies $\epsilon\sim 0.3$ would correspondingly imply rapidly spinning behaviors in the typical supermassive BH population. Unfortunately, this argument could not be considered as conclusive given the present uncertainties inherent to the local mass function determination. 

The parameters $\lambda_0$ and $\xi$ determining the local normalization and redshift evolution of the mean Eddington ratio are also rather well determined by our analysis. Here the relevant point is the mapping between the AGN luminosity function and the BH mass function via the Eddington ratio distribution that enters the continuity equation approach. Specifically,  
lower average values of $\lambda$ tend to originate a flatter
BH mass function at the high-mass end, since large BH masses correspond to moderate AGN luminosities (approximately
$L_{\rm AGN} \propto \lambda\, M_\bullet$ holds) falling in the flatter portion of the AGN luminosity function. Thus local BH mass function determinations with a flatter (steeper) high mass end prefer lower (higher) values of the average $\lambda_0$. This explain the rising trend of the $\lambda_0$ estimate from the BH mass function determination based on the $M_\bullet-M_\star$ relationship, which is the flatter, toward that based on the debiased $M_\bullet-M_\star$ relationship, which features the steeper shape.

On the other hand, the parameter $\xi$ ruling the redshift evolution of the mean Eddington ratio is strongly anti-correlated with $\lambda_0$, so that lower values of $\lambda_0$ correspond to larger values of $\xi$ and hence to stronger redshift evolution. This occurs because moving toward high redshift a progressively larger number of BHs are expected to be active (i.e., the duty cycle typically increases), so that the AGN luminosity function is more directly mapped into the relic BH mass function. In these conditions, values of $\lambda$ substantially exceeding unity are not acceptable, since they would map high AGN luminosities and thus relatively small AGN number density to low BH masses; this will not allow an efficient buildup of the BH mass function at the low mass end.

The parameter $\eta$ dictating the mass dependence of the binary merger rate of supermassive BHs is loosely constrained, though there is a tentative preference for positive values. This is mainly because of two reasons. On the one hand, as to the BH mass function, binary mergers tend to be appreciably relevant only at relatively low redshift $z\lesssim 1$ and at the very massive end $M_\bullet\gtrsim 10^9\, M_\odot$; in fact, outside these ranges in redshift and BH mass, accretion is the main driver for the buildup of the mass function.
On the other hand, as to the stochastic GW background, Equation (\ref{eq|GWstrain}) shows that the parameter $\eta$ appears in an integrated way
as a correction to strong mass dependencies $M_\bullet^{5/3}$ associated to the characteristic GW strain. Therefore it is somewhat expected that the mass dependence in the merger rate can be only loosely constrained. However, it is somewhat interesting to note that bestfit values around $\eta\sim 0.5$ as found here would imply an increased merger rate for more massive BHs; this could reflect an intrinsic physical mechanism favoring them or a selection bias associated to the current nano-Hz GW background data.

The normalization scaling factor $f_{\bullet\bullet\rightarrow \bullet}$ of the supermassive BH merger rate with respect to the galaxy merger rate takes on median values $\lesssim 0.1$. These estimates are rather uncertain but a robust upper limit $\lesssim 1$ at $\gtrsim 2\sigma$ can be derived. The reason is the following. Given the observed bolometric AGN luminosity functions, accretion is able to reproduce by itself the BH demographics for reasonable values of the radiative efficiency and of the mean Eddington ratio. Thus the only solid constraint that our analysis can put on $f_{\bullet\bullet\rightarrow \bullet}$ is that it must not exceed appreciably unity; otherwise, given that the action of mergers is to redistribute masses from the low to the high mass of the mass function, the latter will be flattened and the former lowered too much, in discrepancy with whatever local BH mass function determination.

Finally, the scatter in the $M_\bullet-M_{\rm H}$ relationship inferred from abundance matching is robustly constrained in our analysis by the requirement to reproduce clustering measurements, i.e., the local bias as a function of BH mass. BH mass function determinations based on the observed $M_\bullet-M_\star$ or $M_\bullet-\sigma$ relationship call for considerable values of the scatter $\sigma_{\log M_{\rm H}}\gtrsim 0.5$ dex between BH and halo mass; on the other hand, the BH mass function determination based on the debiased $M_\bullet-\sigma$ relation requires appreciably lower values $\sigma_{\log M_{\rm H}}\gtrsim 0.3$ dex. This is because scaling relations of BH to galaxy properties with lower normalization tend to map on the average a given BH mass into a larger halo mass, hence would imply a higher bias; as a consequence, the scatter in BH to halo mass cannot be too large otherwise the predicted clustering signal would exceed the observational constraints.

\subsubsection{BH demographics and clustering}

In Figure \ref{fig|BHMF} we illustrate the BH mass function resulting from our semi-empirical framework. Solid colored lines show the bestfits to the three local BH mass function determinations (colored symbols) based on the observed $M_\bullet-M_\star$ (red), observed $M_\bullet-\sigma$ (blue) and debiased $M_\bullet-\sigma$ (green) relationship. In addition, the magenta line is the bestfit to the $M_\bullet-M_\star$ determination when a narrow prior on the scaling parameter of the BH to galaxy merger rate $f_{\bullet\bullet-\rightarrow \bullet}$ around unity is adopted. All the fits are remarkably good, highlighting that the parameterizations adopted in our semi-empirical framework are sensible and sufficient to describe the BH demographics. Moreover, the dashed colored lines illustrate the corresponding active BH mass function at high redshift $z\sim 5$, which should be compared with the observational estimates by \cite{He2024,Lai2024,Taylor2025,Matthee2024,Kokorev2024,Wu2022} at $z\sim 4-6$ (dark grey symbols). Note that the high-redshift active BH mass function has not been included in the fitting process, so the good agreement found with these independent data is a genuine prediction of our model, and further corroborates our choice of a redshift dependent mean Eddington ratio.

\begin{figure}[t!]
    \centering
    \includegraphics[width=1.\textwidth]{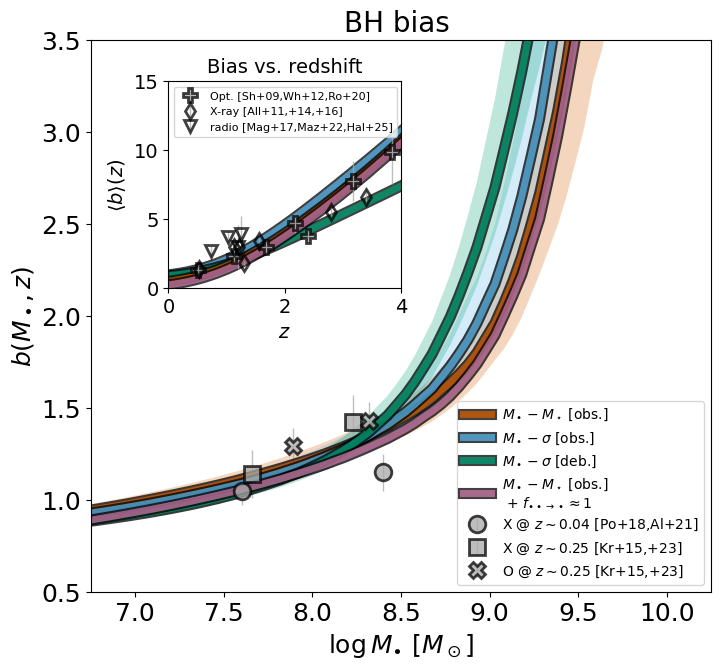}
    \caption{Large-scale bias as a function of BH mass. Data are from X-ray AGNs at $z\sim 0.04$ by \cite{Powell2018,Allevato2021} (circles), from X-ray (squares) and optical (crosses) AGNs at $z\sim 0.25$ by \cite{Krumpe2015,Krumpe2023}. Colored solid lines with shaded areas illustrate the outcomes (median and $2\sigma$ credible intervals) from our semi-empirical framework model and MCMC analysis, with the same color code as in previous Figures. The inset shows the BH bias as a function of redshift, averaged over the supermassive BH mass function. Colored lines are the outcomes from our semi-empirical framework, while symbols report observational estimates from optical (plus signs; \cite{Shen2009,White2012,Ross2020}), X-ray (diamonds; \cite{Allevato2011,Allevato2014,Allevato2016}) and radio-selected (inverse triangles; \cite{Magliocchetti2017,Mazumder2022,Hale2025}) samples. Note that the redshift dependence has not been fitted upon, and constitutes just an a-posteriori validation of our semi-empirical framework.}
    \label{fig|BHbias}
\end{figure}

In Figure \ref{fig|BHbias} we compare the large-scale bias $b(M_\bullet,z)$ as a function of BH mass from our semi-empirical framework for the different local BH mass function determinations (colored lines) to clustering data at $z\sim 0.01-0.25$ by \cite{Powell2018,Allevato2021,Krumpe2015,Krumpe2023}, finding an excellent agreement. In addition, the inset shows the BH bias as a function of redshift, after averaging over the supermassive
BH mass function, i.e. computing $\langle b\rangle(z) \equiv \int{\rm d}\log M_\bullet\, N(\log M_\bullet,z)\, b(M_\bullet,z)/$ $\int{\rm d}\log M_\bullet\, N(\log M_\bullet,z)$. Although this redshift dependence of the bias has not been fitted upon, the outcomes from our semi-empirical framework agree reasonably well
with the observational estimates from optical (plus signs; \cite{Shen2009,White2012,Ross2020}), X-ray (diamonds; \cite{Allevato2011,Allevato2014,Allevato2016}) and radio-selected (inverse triangles; \cite{Magliocchetti2017,Mazumder2022,Hale2025}) samples.

\begin{figure}[t!]
    \centering
    \includegraphics[width=1.\textwidth]{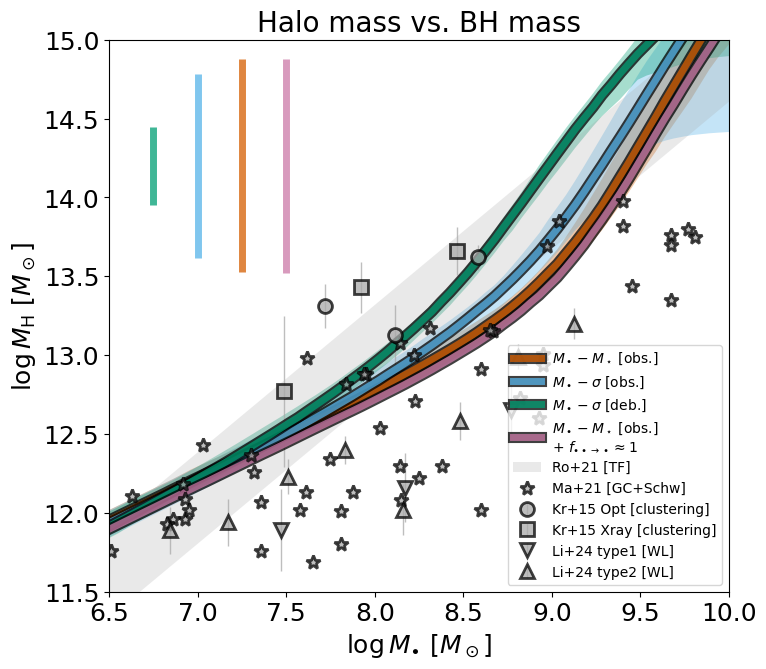}
    \caption{Halo mass vs. BH mass relationship. Data symbols refer to estimate from clustering of X-ray (squares) and optical (circles) AGNs by \cite{Krumpe2015}, from weak lensing measurements of type-1 (inverse triangles) and type-2 (triangles) AGNs, from globular cluster dynamics (stars) by \cite{Marasco2021}, and from Tully-Fisher kinematics for reverberation-mapped AGNs by \cite{Robinson2021} (grey shaded area).  Colored solid lines with shaded areas illustrate the outcomes (median and $2\sigma$ credible intervals) from our semi-empirical framework and MCMC analysis on BH demographics and clustering, with the same color code as in previous Figures; note that the data reported here have not been fitted upon, and constitute just an a-posteriori validation. In the top left corner the colored errorbars represent the bestfit scatter $\sigma_{\log M_{\rm H}}$ on the BH to halo mass relationship inferred from our analysis.}
    \label{fig|BHabma}
\end{figure}

In Figure \ref{fig|BHabma} the related local BH to halo mass relationship is displayed, and compared with various observational estimates by \cite{Marasco2021,Krumpe2015,Li2024,Robinson2021} derived by a variety of methods (e.g., clustering, kinematics, weak lensing). Actually the data are rather dispersed, partly because of intrinsic variance in the relationships, and partly because of possible observational systematics. In particular, there is a tendency for weak lensing estimates to predict a BH to halo mass relation with a lower normalization with respect to clustering measurements. 
Note that in view of these observational uncertainties, the BH to halo mass relation has not been fitted upon in our analysis. However, the relation derived via abundance matching from our semi-empirical framework, also in view  of the estimated intrinsic scatter $\sigma_{\log M_{\rm H}}\gtrsim 0.3-0.5$ dex (colored errorbars in the plot), is consistent with the cloud of datapoints.

\begin{figure}[t!]
    \centering
    \includegraphics[width=1.\textwidth]{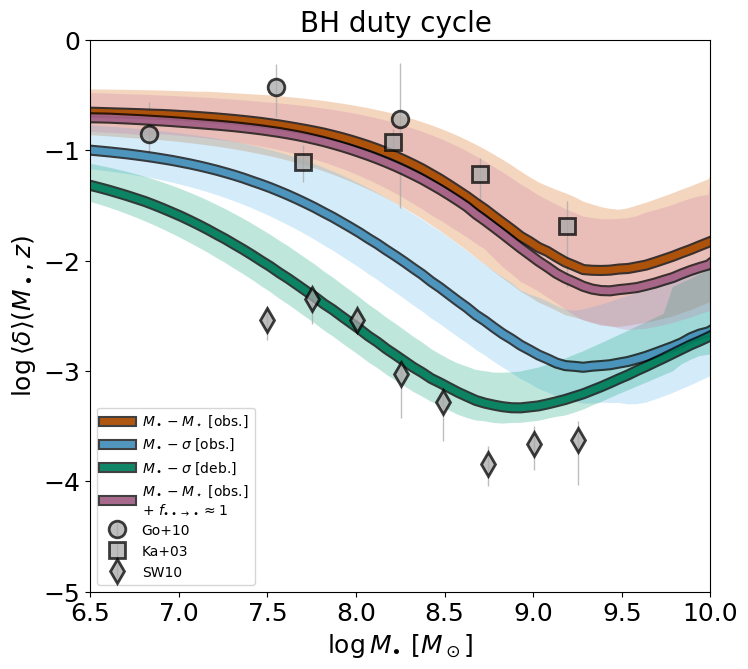}
    \caption{Local duty cycle as a function of BH mass. Data are from AGN active fractions by  \cite{Goulding2010} (circles), \cite{Kauffmann2003} (squares) and \cite{Schulze2010} (diamonds). Colored solid lines with shaded areas illustrate the outcomes (median and $2\sigma$ credible intervals) from our semi-empirical framework and MCMC analysis, with the same color code as in previous Figures; note that the data reported here have not been fitted upon, and constitute just an a-posteriori validation.}
    \label{fig|BHduty}
\end{figure}

Figure \ref{fig|BHduty} illustrates the local BH duty cycle as a function of BH mass from our semi-empirical framework for the different local BH mass function determinations (colored lines), compared with classic observational estimates by \cite{Kauffmann2003,Goulding2010,Schulze2010} derived from AGN active fractions in the local Universe. First of all, it is important to notice that there is a rather large dispersion in the data, which partly reflects differences in bolometric corrections, luminosity thresholds, and systematic offsets in the BH mass estimates; for these reasons the duty cycle has not been fitted upon in our analysis. Nevertheless, the duty cycle reconstructed from the continuity equation is found to be broadly consistent with these observational estimates. More in detail, higher duty cycle around $1-10\%$ are produced for the models fitting the local BH mass function determination based on the observed $M_\bullet-M_\star$ relation, while smaller duty cycles $\sim 0.1-1\%$ are originated in the model fitting BH mass function based on the debiased $M_\bullet-\sigma$ relationship. 

\begin{figure}[t!]
    \centering
    \includegraphics[width=.75\textwidth]{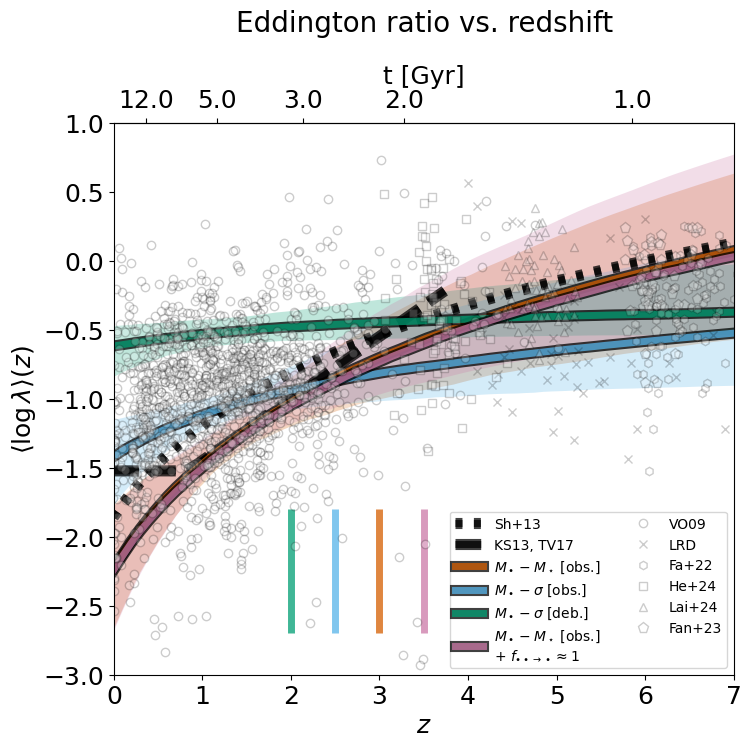}
    \includegraphics[width=0.9\textwidth]{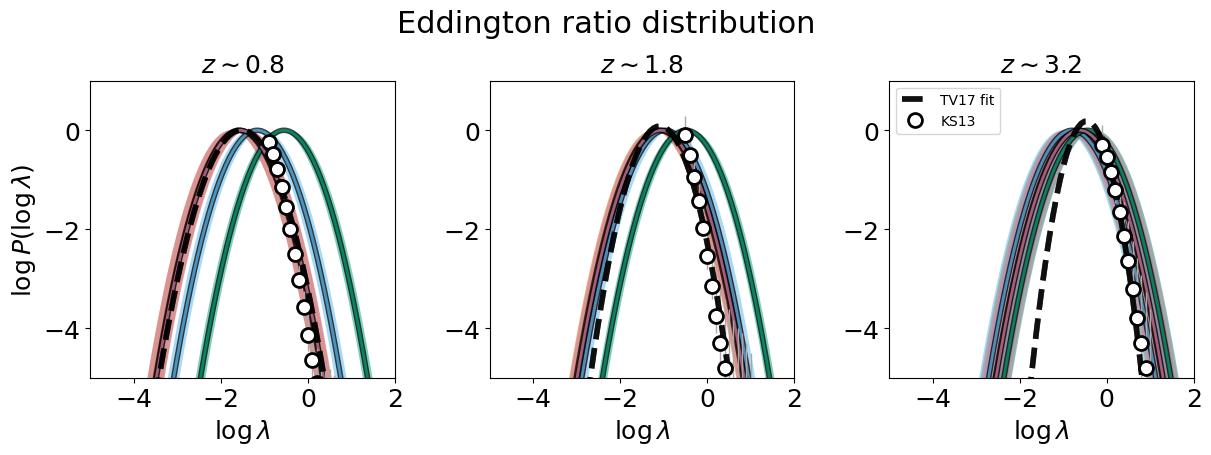}
    \caption{Top panel: Average Eddington ratio vs. redshift (bottom axis) or cosmic time (top axis). Data are from: classic measurements from X-ray/mid-IR/optical AGNs by \cite{Vestergaard2009,Nobuta2012,Dai2014,Duras2020} (circles),  broad-absorption-line quasars by \cite{He2024} (squares), \cite{Lai2024} (triangles) \cite{Farina2022} (diamonds) and \cite{Fan2023} (pentagons); little red dots by \cite{Harikane2023,Greene2024,Matthee2024,Maiolino2024} (crosses). 
    For reference, the dashed line illustrates the mean Eddington ratio found for type-1 AGNs at $z\lesssim 4$ by \cite{Kelly2013}, as fitted by \cite{Tucci2017}; the dotted line is the dependence assumed on phenomenological grounds by \cite{Shankar2013}. Colored solid lines with shaded areas illustrate the outcomes (median and $2\sigma$ credible intervals) from our semi-empirical framework and MCMC analysis, with the same color code as in previous Figures; note that the data reported here have not been fitted upon, and constitute just an a-posteriori validation. The colored errorbars in the bottom left corner represent the log-normal scatter in the Eddington ratio distribution adopted in our model. Bottom panels: Eddington ratio distribution at three representative redshifts $z\approx 0.8$ (left) $1.8$ (middle) and $3.2$ (right). Circles illustrate the estimates for type-1 AGNs by \cite{Kelly2013}, 
    plotted for the range in $\log \lambda$ where the completeness for the flux-limited SDSS sample is above 10\%. Dashed line is the corresponding lognormal fit by \cite{Tucci2017}. Colored lines with shaded areas show the outcomes of our semi-empirical framework, as above.}
    \label{fig|lamdaz}
\end{figure}

In the top panel of Figure \ref{fig|lamdaz} we showcase the mean Eddington ratio as a function of redshift as derived from our analysis for the different local BH mass function determinations (colored lines). These outcomes are compared with a multitude of observational estimates from X-ray/mid-IR/optical AGNs by \cite{Vestergaard2009,Nobuta2012,Dai2014,Duras2020}, broad-absorption-line quasars by \cite{He2024,Lai2024,Farina2022,Fan2023}, and little red dots by \cite{Harikane2023,Greene2024,Matthee2024,Maiolino2024}. The various datasets are somewhat heterogeneous, so that the comparison with the model outcomes should be taken only as indicative. In fact, we stress that the parameters $\lambda_0$ and $\xi$ determining in our model the redshift-dependent Eddington ratio after Equation (\ref{eq|lambda}) have not been fitted on the data reported here, but instead have been determined by reproducing the BH demographics and clustering via the continuity equation approach. Nevertheless, it is reassuring that the model outputs on $\lambda(z)$ are broadly consistent with these observational determinations for individual AGNs.
As already mentioned, the outcome referring to the BH mass function determination based on the observed $M_\bullet-M_\star$ relation tends to predict rather low mean Eddington ratios $\lambda\lesssim 0.01$ in the local Universe and a steep rise toward high redshift values $\lambda\sim 1$ at $z\sim 6$. This behavior agrees well with the average Eddington ratio observationally determined for type-1 AGNs up to $z\lesssim 4$ by \cite{Kelly2013}, as fitted by \cite{Tucci2017}; it also accords with the dependence assumed on phenomenological grounds by \cite{Shankar2013}. On the other hand, the model outcomes referring to the BH mass function determinations based on the observed or debiased $M_\bullet-\sigma$ relations point toward somewhat higher current normalization $\lambda\lesssim 0.3$ and an appreciably weaker evolution toward high redshift. This behavior reflects the well-known degeneracy between the local duty cycle and the evolution of the Eddington ratio (see \cite{Shankar2013}). In the bottom panel of Figure \ref{fig|lamdaz} we illustrate the full Eddington ratio distribution at three representative redshifts $z\approx 0.8$ (left) $1.8$ (middle) and $3.2$ (right). The outcomes of our semi-empirical framework are in good accord with the estimates for type-1 AGNs by \cite{Kelly2013} (circles) and with the corresponding lognormal fits by \cite{Tucci2017} (dashed lines).

\begin{figure}[t!]
    \centering
    \includegraphics[width=1.\textwidth]{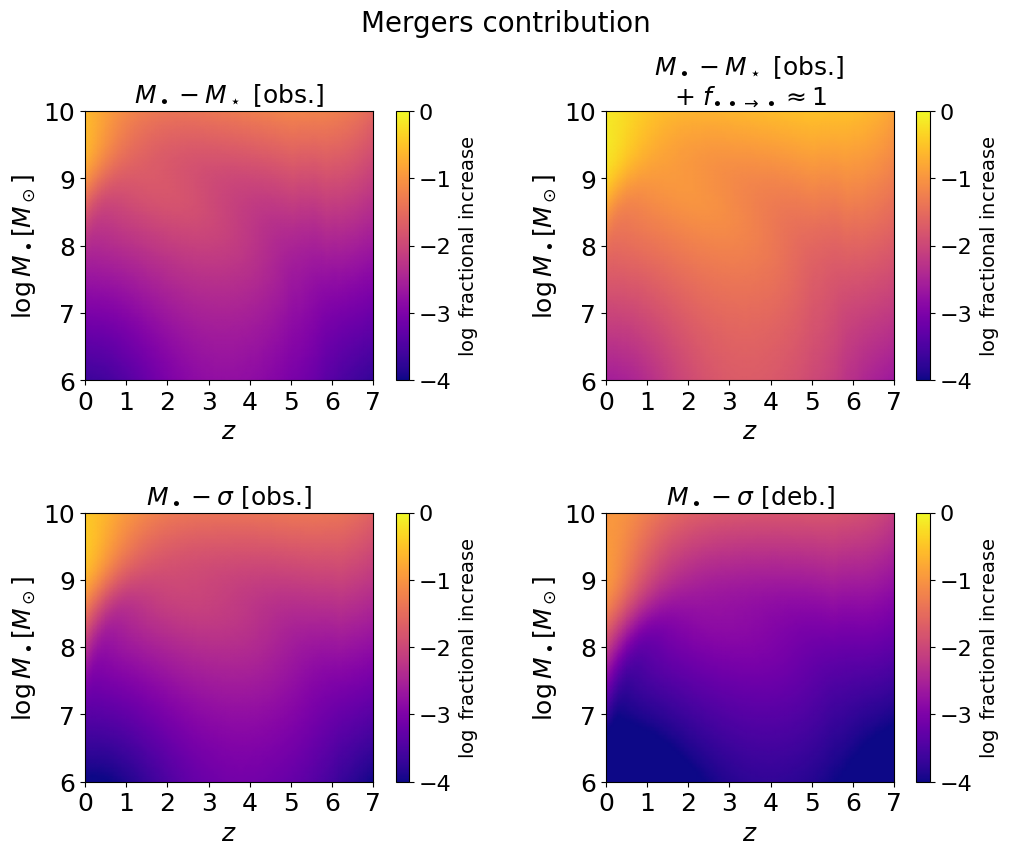}
    \caption{The relative contribution of binary BH mergers to the increase of the BH mass function in different mass and redshift bins (color-coded). Panels refer to the analysis with different determinations of the local BH mass function: top left is for SMF plus observed $M_\bullet-M_\star$; top right is for SMF plus observed $M_\bullet-M_\star$ plus a narrow prior centered on $f_{\bullet\bullet\rightarrow \bullet}\approx 1$; bottom left is for VDF plus observed $M_\bullet-\sigma$; bottom right is for VDF plus debiased $M_\bullet-\sigma$.}
    \label{fig|relative}
\end{figure}

In Figure \ref{fig|relative} we illustrate the relative contributions of mergers to the increase in the BH mass function, sliced in different bins of BH mass and redshift. The color scale illustrates the ratio between the production rate of the mass function due to mergers (i.e., the positive term in the Smoluchowski equation) normalized to the sum of this quantity and of the analogous one due to accretion. It is evident that the main impact of mergers in increasing the mass function occurs toward low redshift $z\lesssim 1.5$ and for relatively high masses $M_\bullet \gtrsim$ some $10^8\, M_\odot$. Such a contribution
is extended somewhat toward higher redshift and smaller BH masses for the BH mass function determination based on the $M_\bullet-M_\star$ relation, which features the highest normalization at the high mass end, and thus leave more rooms for redistribution of the masses there.
An additional yet modest contribution occurs at intermediate redshift $z\sim 2-4$ because the redshift increase in the BH (and galaxy) merger rate cooperates with a still appreciably high BH mass function; at higher $z$ the effect is less evident since the merger rate stays put but the BH mass function drops considerably. Finally, it is worth noticing that for very large BH masses $M_\bullet \gtrsim 10^9\, M_\odot$ mergers are quite relevant even at substantial redshifts $z\gtrsim 6$, though clearly such over-massive BHs in the primordial Universe would be rare.

\subsubsection{Stochastic GW background}

\begin{figure}[t!]
    \centering
    \includegraphics[width=1.\textwidth]{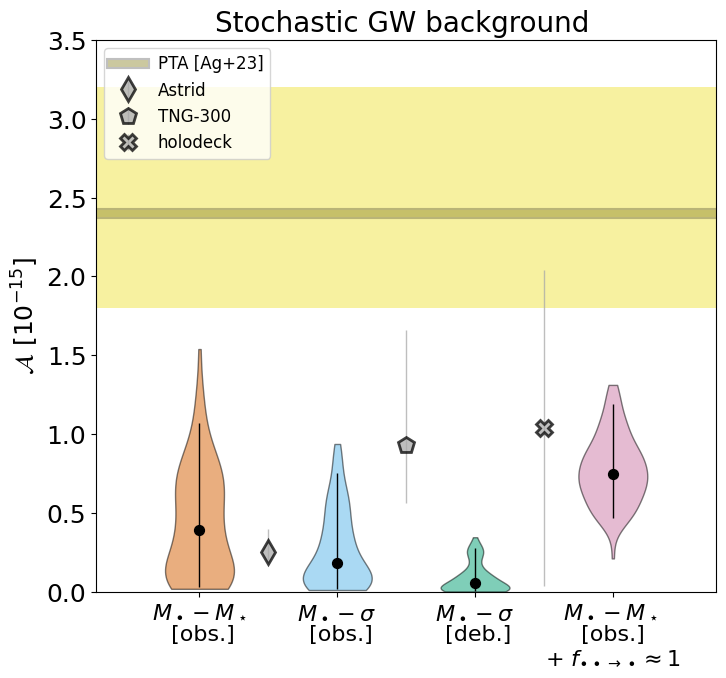}
    \caption{Normalized stretch amplitude of the gravitational wave background. Colored violins illustrate the full posterior distributions from our semi-empirical framework and MCMC analysis, with the same color code as in previous Figures (and also specified on the bottom axis); the black dots and solid lines illustrate the median and the $2\sigma$ credible interval. Grey symbols display the results from the \texttt{Astrid} (rhomboid; \cite{Chen2025}) and \texttt{TNG-300} numerical simulations (pentagon; \cite{Chen2025}), and from the \texttt{holodeck} suite (cross; \cite{Agazie2023}). The measurement by the \texttt{NANOGrav} collaboration \cite{Agazie2023} is displayed by the yellow solid line with shaded area (median and $90\%$ credible interval).}
    \label{fig|PTA}
\end{figure}

In Figure \ref{fig|PTA} we illustrate our posterior estimates for the normalized stretch amplitude of the nano-Hz GW background for the different local BH mass function determinations (colored violin plots), 
with the black dots and solid errorbars illustrating the median and the $2\sigma$ credible intervals. These should be compared  with the observational estimate of the background amplitude by the \texttt{NANOGrav} collaboration \cite{Agazie2023} (yellow solid line with shaded area). It appears evident that the outcomes of our analysis do not favor a substantial contribution to the observed GW background from supermassive BH mergers. Even in the extreme case (magenta violin plot) where we have adopted the BH determination based on the $M_\bullet-M_\star$ relation (which has the highest normalization) and we have imposed a narrow prior around unity on the rescaling parameter $f_{\bullet\bullet\rightarrow\bullet}$ between the BH and the galaxy merger rate, our analysis predicts that only a fraction $\lesssim 30-50\%$ (at $\sim 3\sigma$) of the observed background can be traced back to supermassive BH mergers. Our results are in good agreement with the outcomes of the \texttt{Astrid} and \texttt{TNG-300} numerical simulations (see \cite{Chen2025}) and of the \texttt{holodeck} suite (see \cite{Agazie2023}); in fact, the latter two have a median closer to our maximized result with $f_{\bullet\bullet\rightarrow\bullet}\approx 1$, but the large uncertainties make them consistent also with our more conservative estimates.

\begin{figure}[t!]
    \centering
    \includegraphics[width=1.\textwidth]{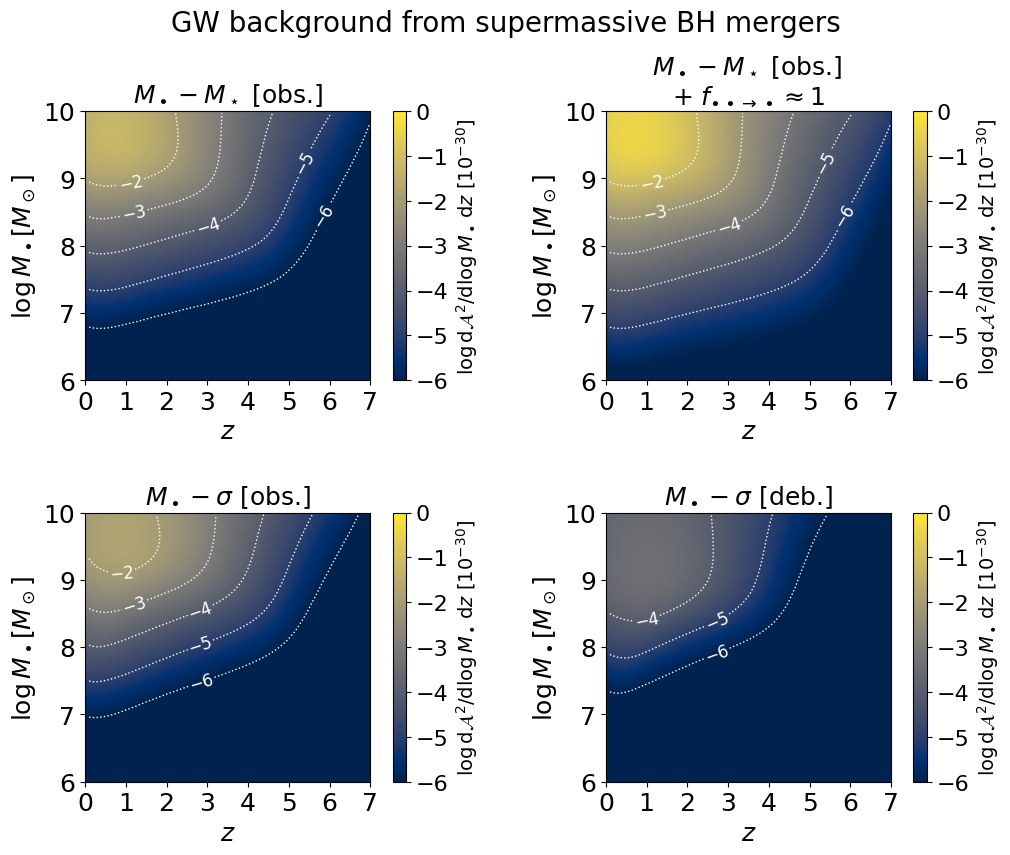}
    \caption{The contribution to the GW stochastic background from supermassive BH mergers sliced in redshift and BH mass. Panels refer to the analysis with different determination of the local BH mass function: top left is for SMF plus observed $M_\bullet-M_\star$; top right is for SMF plus observed $M_\bullet-M_\star$ plus a narrow prior centered on $f_{\bullet\bullet\rightarrow \bullet}\approx 1$; bottom left is for VDF plus observed $M_\bullet-\sigma$; bottom right is for VDF plus debiased $M_\bullet-\sigma$. The color-scale and white contours display logarithmic values of the normalized (squared) strain parameters $\mathcal{A}^2$ in bins of BH mass $\log M_{\bullet}$ and redshift $z$.}
    \label{fig|GWback}
\end{figure}

In Figure \ref{fig|GWback} we highlight the contribution of supermassive BH mergers to the nano-Hz GW stochastic background sliced in redshift and BH mass, for the different local BH mass function determinations (different panels).
Most of background is contributed by relatively low redshift $z\lesssim 3$ sources and large BH masses $M_\bullet\gtrsim 10^{9}\, M_\odot$. Such a contribution extends toward higher redshift and smaller BH masses for the BH mass function determination based on the $M_\bullet-M_\star$ relation,  which features the highest normalization at the high mass end; this behavior is then reinforced when the binary BH merger rate is maximized and imposed to follow the galaxy merger rate (i.e., when $f_{\bullet\bullet\rightarrow\bullet}\approx 1$; top right panel). This figure also highlights that the tension between BH demographics and GW background is robust to residual uncertainties on the basic inputs of our data-driven model. For example, one could implement variations in the AGN luminosity function at $z\gtrsim 4.5$ (see Appendix \ref{app|AGN_LF}), which may account for a steeper trend/higher normalization  with respect to \cite{Shen2020} as claimed by \cite{Giallongo2019,Kulkarni2019,Barlow2023,Barlow2025} or possibly suggested by an increased UV space density due to little red dots in case these are truly AGN dominated (see \cite{Kokorev2024}). However, the impact on the GW background from such changes at relatively high redshift and small to moderate AGN luminosities/BH masses would be minor (more quantitative analysis is presented in Appendix \ref{app|tests}).

A possible tension between the GW background expected from BH mergers and the one measured by PTA has been noted and discussed before in the literature (e.g., \cite{Sesana2016,Chen2019,Agazie2023,Agazie2023b,Middleton2021,Villalba2022,SatoPolito2025,Casey2025,Chen2025,Kis2025}). However, these studies lack a self-consistent treatment of BH mergers, in that the related rates (usually derived from galaxy merger rates via a BH-galaxy scaling relation) are used to compute the expected GW background, but are not employed at the same time to build up and check the related impact on the mass function. Our analysis point out the tension via a semi-empirical, data-driven model based on the continuity plus Smoluchowski equation, that treats unitarily accretion and mergers, incorporates the latest observational determinations of the AGN luminosity functions and galaxy merger rates, and is jointly constrained by local BH demographics and GW background.

\begin{figure}[t!]
    \centering
    \includegraphics[width=1.\textwidth]{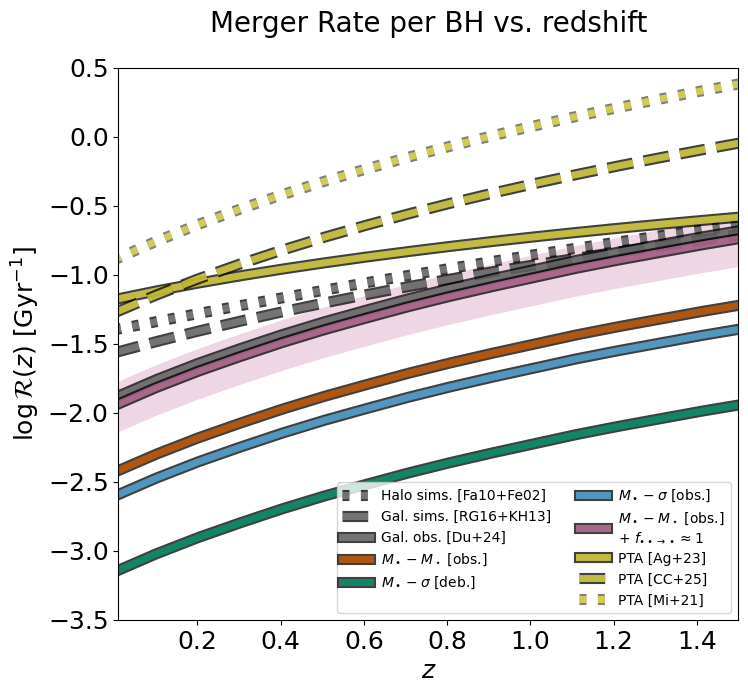}
    \caption{The merger rate per descendant BH vs. redshift, referred to a reference BH mass $M_\bullet\sim 10^{10}\, M_\odot$ as a representative contributor to the GW background. Colored solid lines illustrate the outcomes from our semi-empirical framework and MCMC analysis, with the same color code as in previous Figures: the median from the posterior is reported, but the uncertainty for the red, blue and green cases essentially extends from the upper bound of the magenta line downwards (i.e., our model can only set upper limits on $f_{\bullet\bullet\rightarrow\bullet}$, see Fig. \ref{fig|MCMC}).     
    The yellow lines are the reconstructed merger rate per BH from various analysis of PTA data: solid refers to \cite{Agazie2023b}, dashed to \cite{Casey2025}, and dotted to \cite{Middleton2021}. The grey lines are illustrated for comparison: solid is the major merger rate per galaxy estimated from \texttt{JWST} galaxy pairs counts by \cite{Duan2025}; dashed is the major merger rate per galaxy from the \texttt{Illustris} simulation by \cite{Rodriguez2015}, converted to the reference BH mass via the BH-stellar mass relation by \cite{Kormendy2013}; dotted is the major merger rate per dark matter halo by \cite{Fakhouri2010}, converted to the reference BH mass via the BH-halo mass relation by \cite{Ferrarese2002}.}
    \label{fig|Rmergz}
\end{figure}

The situation is better clarified in Figure \ref{fig|Rmergz}, where we illustrate the merger rate per descendant BH (basically Equation \ref{eq|mergratebase} integrated over the mass ratio distribution) vs. redshift, referred to a reference BH mass $M_\bullet\sim 10^{10}\, M_\odot$ which is taken as a representative contributor to the GW background. Note that the mass dependence is anyway not an issue since it is weak and the range of BH masses contributing to the background is rather limited (see above). Colored solid lines illustrate the median outcomes from our semi-empirical framework and MCMC analysis, for the different local BH mass function determinations. Essentially they are rescaled version (by the parameter $f_{\bullet\bullet\rightarrow\bullet}$) of the solid gray curve, which represents the galaxy merging rate as estimated from pairs counts by \texttt{JWST}.  However, note that the $2\sigma$ uncertainty (not reported for the sake of clarity otherwise the plot would look crowded) for the red, blue and green cases essentially extends from the magenta line downwards, since our model can only set an upper limit to the merging rates per BH, as it is evident from the posterior of the parameter $f_{\bullet\bullet\rightarrow\bullet}$ in Fig. \ref{fig|MCMC}. For comparison, the dashed gray line refers to the major merger rate per galaxy from the \texttt{Illustris} simulation by \cite{Rodriguez2015}, converted to the reference BH mass via the BH-stellar mass relation by \cite{Kormendy2013}, while the dotted gray line is the major merger rate per dark matter halo by \cite{Fakhouri2010}, converted to the reference BH mass via the BH-halo mass relation by \cite{Ferrarese2002}. 

The take-home message is that to reproduce the high mass end of the estimated local mass functions (regardless of the specific determination), our model restricts the merger rate to be smaller or in any case not to exceed that inferred from galaxy pairs by \cite{Duan2025,Puskas2025}. This is in turn slightly lower at $z\lesssim 0.5$ than that measured for galaxies in hydrodynamic simulations and for halos in $N-$body codes. Contrariwise, fitting the GW background measured by PTA experiments would require the merger rate per BH to be appreciably higher, as illustrated by the yellow curves that refer to various parametric analyses of the PTA data by \cite{Middleton2021,Agazie2023b,Casey2025}. However, these high rates would inevitably corrupt the local BH mass function at the high mass end, in a way that cannot be effectively cured by any reasonable accretion-related parameter set. Moreover, these merger rates per BH required to fit the PTA data turn out to be appreciably higher than the merger rate per galaxies or per halo expected from observations of galaxy pairs and from numerical simulations, an occurrence which would be in itself rather problematic to explain.

In principle, one could envisage physically-motivated modifications of the current setup to potentially ease the tension between the PTA signal and the model predictions. For example, a local BH mass function more skewed toward larger masses than the ones presented in this work could provide room for more frequent and/or more massive BH mergers, thus boosting the implied GW background signal. This could originate from a particularly flat stellar mass function at the high-mass end, and/or from an appreciably high normalization of the $M_\bullet-\sigma$ relationship (as may be the case for cored ellipticals, see \cite{Saglia2024}; also K. M. Varadarajan in preparation). Another possibility may involve a radiative efficiency increasing with the BH mass (see \cite{Cao2008,Shankar2013}) which progressively reduces the number density of the most massive BHs, so as to reconcile the local BH mass function with the one derived from accretion even in the event of an increased number of BH mergers. However, a couple of reasons suggest that these attempts are likely doomed to fail. First, the merger rate density of galaxies is nowadays well constrained by \texttt{JWST} data, and even requiring the merger rate density of BHs to mirror it is not sufficient to reproduce the PTA background, as shown by our analysis. Therefore, independently of the shape of the local BH mass function, explaining all the GW background via BH mergers would require to assume that the merger rate per BH is appreciably larger than that of galaxies. Second, leaving more room to mergers at late times in building up the local BH mass function at the high-mass end could be dangerous, since mergers would also tend to reduce the number of low-mass BH (which merge into larger ones) so distorting the mass function there. All in all, solving the tension by acting solely on the accretion-related parameters would require some fine-tuning and would remain a very challenging task. We plan to perform an in-depth analysis of these and further possibilities in a dedicated future work.

\section{Summary}\label{sec|summary}

We have explored the evolution of the supermassive BH population via a semi-empirical approach with minimal assumptions and data-driven inputs. This is based on a continuity plus Smoluchowski equation framework aimed to describe jointly the two main modes of BH growth, namely gas accretion and binary mergers. We have educatedly parameterized the main uncertainties associated to each BH growth channel, and inferred constraints on these in a Bayesian setup from estimates of the local BH mass function, clustering, and measurements of the nano-Hz stochastic GW background. Our main findings are as follows.

\begin{itemize}

\item Different local mass function determinations imply substantially different values for the parameters regulating the accretion mode of the supermassive BH population.
The BH mass function based on the stellar mass function and observed $M_\bullet-M_\star$ relationship requires low radiative efficiency $\epsilon\sim 0.06$, small local average values of the Eddington ratio $\lambda\sim 10^{-2}$ with a rather strong dependence on redshift $\lambda\propto (1+z)^{2.5}$;  a high dispersion $\sigma_{\log M_{\rm H}}\sim 0.7$ in the relationship between BH and host halo mass is also required to reproduce the BH clustering. The BH mass function based on the velocity dispersion function and the observed $M_\bullet-\sigma$ relationship allows for standard values of the efficiency $\epsilon\sim 0.1$, local average Eddington ratios $\lambda\lesssim 10^{-1}$ with a mild redshift evolution $\lambda\propto (1+z)$, and scatter of $\sigma_{\log M_{\rm H}}\sim 0.6$ in the BH-halo mass relation. Finally, the BH mass determination based on the debiased $M_\bullet-\sigma$ relationship requires considerably high values of the efficiency $\epsilon\sim 0.3$, local average Eddington ratio $\lambda\lesssim 0.3$ modestly evolving with redshift $\lambda\propto (1+z)^{0.4}$, and appreciably lower scatter $\sigma_{\log M_{\rm H}}\sim 0.3$ in the BH vs. halo mass relation. The difference in efficiency could have far-reaching implications on the BH spinning state of the BH population, though systematic uncertainties in the local BH mass function determination do not allow to draw firm conclusions yet.

\item In all the above instances the fits from our semi-empirical framework to the local BH mass function and BH clustering are remarkably good. Moreover, the outcomes of our analysis are able to reproduce auxiliary observables that were not included in the fitting procedure: the active BH mass function at high redshift $z\sim 3-6$ comprising the recently discovered population of little red dots, the local BH vs. halo mass relationship, the local BH duty cycle and the evolution of the mean Eddington ratio across cosmic times. Such an agreement supports our semi-empirical, data-driven framework and the outputs of the continuity equation approach. 

\item Regardless of the adopted local BH mass function determination, we have found that the impact of binary BH mergers on the BH demographics is rather limited, and becomes only relevant for BH masses $M_\bullet\gtrsim$ some $10^8\, M_\odot$ at relatively low redshifts $z\lesssim 1.5$. This result stems from 
the fact that most of the local BH mass density is acquired via gas accretion, with mergers playing a secondary role at low redshift and large BH masses.
More quantitatively, the best-fit values for the rescaling factor of the BH to galaxy merger rate is $f_{\bullet\bullet\rightarrow\bullet}\lesssim 10^{-1}$. These estimates are quite uncertain but a robust upper limit $\lesssim 1$ can be derived at $\gtrsim 2\sigma$. 
There is a tentative evidence for a weak direct dependence of the binary BH merger rate on mass $M_\bullet^{\sim 1/2}$, though the uncertainty on the slope is still very large. 

\item Our analysis does not favor a substantial contribution from supermassive BH mergers to the observed GW background. Irrespective of the local BH mass function determination, we have constrained the contribution of binary BH mergers to be less than $\lesssim 30\%$ (at $\sim 3\sigma$) of the measured background. Even in the extreme case where we adopt the BH determination with the highest normalization at the massive end (the one based on the $M_\bullet-M_\star$ relation) and a BH merger rate strictly mirroring the galactic one (i.e. $f_{\bullet\bullet\rightarrow\bullet}\approx 1$), we have bound the contribution from BH mergers to the GW background at around $\lesssim 50\%$ at $\sim 3\sigma$. Finally, we have highlighted that binary BH mergers at relatively low redshift $z\lesssim 3$ with large BH masses $M_\bullet\gtrsim 10^{9}\, M_\odot$ are the main contributors to the observed GW background. 

\end{itemize}

In a future perspective, it would be beneficial to input our semi-empirical framework with improved determinations of the AGN luminosity functions and active BH mass functions, especially toward high redshift. In this respects, new data from \texttt{Euclid}, \texttt{JWST}, \texttt{e-ROSITA} and eventually \texttt{NewATHENA} will clear residual uncertainties on bolometric corrections, obscured AGN fraction, and abundance of BHs with faint bolometric luminosities. In addition, it would be extremely important to solve the longstanding debate concerning the local BH mass function determination. Although to build a BH mass function based on an extended sample of quiescent BHs with robustly determined masses would remain challenging in the near future, a great step forward could be to identify the most fundamental BH-galaxy relationship and assess the observational and systematics issues that may affect and bias it. Wide galaxy surveys that will be conducted by the \texttt{Rubin-LSST} and eventually by \texttt{ELT} will help in this respect. All the above will improve appreciably the constraints on the parameters regulating the accretion mode for the growth of the BH population, like the AGN luminosity function, the radiative efficiency (hence BH spins), and the redshift-dependent Eddington ratio distributions. 

As for the impact of binary BH mergers on the growth of the supermassive BH population, in the future it will be effectively constrained by more precise measurements of the nano-Hz GW background from PTA experiments; this will be possible thanks to the advent of the \texttt{SKAO} and its precursors, that will allow the monitoring of a substantially increased amounts of pulsars. These data will greatly improve the constraints on the mass and possibly the redshift dependence of the BH merger rate, allowing to confirm the conclusion of the present analysis concerning a modest contribution of binary BH mergers to the GW background. If so, the implications would be enormous. On the one hand, this would be a good news for \texttt{LISA}, since  supermassive BH mergers may originate an appreciable background noise in the mHz band, degrading the effective sensitivity of the antennae to many primary science goals of the mission (e.g., \cite{Huang2024}), like the detection of individual massive BH binaries, verified galactic binaries, and extreme mass ratio inspirals. On the other hand, our current results could motivate an exciting search for alternative astrophysical sources or even rise the hype for a primordial, cosmological origin of the remaining signal measured by PTA experiments.

\acknowledgments

We thank the anonymous referee for constructive and helpful comments.
This work was partially funded from the projects: INAF GO-GTO Normal 2023 funding scheme with the project "Serendipitous H-ATLAS-fields Observations of Radio Extragalactic Sources (SHORES)"; INAF Large Grant 2022 project "MeerKAT and LOFAR Team up: a Unique Radio Window on Galaxy/AGN co-Evolution; INAF Large GO 2024 project "MeerKAT and Euclid Team up: Exploring the galaxy-halo connection at cosmic noon"; ``Data Science methods for MultiMessenger Astrophysics \& Multi-Survey Cosmology'' funded by the Italian Ministry of University and Research, Programmazione triennale 2021/2023 (DM n.2503 dd. 9 December 2019), Programma Congiunto Scuole; Italian Research Center on High Performance Computing Big Data and Quantum Computing (ICSC), project funded by European Union - NextGenerationEU - and National Recovery and Resilience Plan (NRRP) - Mission 4 Component 2 within the activities of Spoke 3 (Astrophysics and Cosmos Observations); European Union — NextGenerationEU under the PRIN MUR 2022 project n. 20224JR28W “Charting unexplored avenues in Dark Matter”. MB acknowledges that this article was produced while attending the PhD program in PhD in Space Science and Technology at the University of Trento, Cycle XXXIX, with the support of a scholarship financed by the Ministerial Decree no. 118 of 2nd March 2023, based on the NRRP - funded by the European Union - NextGenerationEU - Mission 4 "Education and Research", Component 1 "Enhancement of the offer of educational services: from nurseries to universities” - Investment 4.1 “Extension of the number of research doctorates and innovative doctorates for public administration and cultural heritage” - CUP E66E23000110001. LB acknowledges financial support from the German Excellence Strategy via the Heidelberg Cluster of Excellence (EXC 2181 - 390900948) STRUCTURES. HF acknowledges support at Fudan University from the Shanghai Super Post-doctoral Excellence Program grant No. 2024008. DR acknowledges support from the University of Southampton via the Mayflower studentship.

\begin{appendix}

\section{The high-redshift AGN luminosity function}\label{app|AGN_LF}

In the main text we have adopted as our fiducial input the AGN luminosity function by \cite{Shen2020}, and in particular their fitting formula reported in Equation (\ref{eq|shen}). This currently constitutes the best homogenized rendition of different multi-wavelength datasets on the AGN luminosity function, converted to bolometric and accounting for obscured fraction with common corrections. However, recently the shape of the \cite{Shen2020} luminosity function at $z\gtrsim 4.5$ has been somewhat challenged by new estimates from optical and X-ray data (see \cite{Wolf2021,Greene2024,Maiolino2024,Barlow2023,Barlow2025}). The issue has been also reinforced by the population of overmassive, X-ray underluminous BHs recently discovered by \texttt{JWST} (e.g., \cite{Ubler2023,Furtak2023,Harikane2023,Kokorev2023,Stone2024,Maiolino2024jades}), partly located in the so called `little red dots' sources (see \cite{Kocevski2023,Matthee2024,He2024,Lai2024,Kokorev2024,Akins2025}).

\begin{figure}[t!]
    \centering
    \includegraphics[width=1.\textwidth]{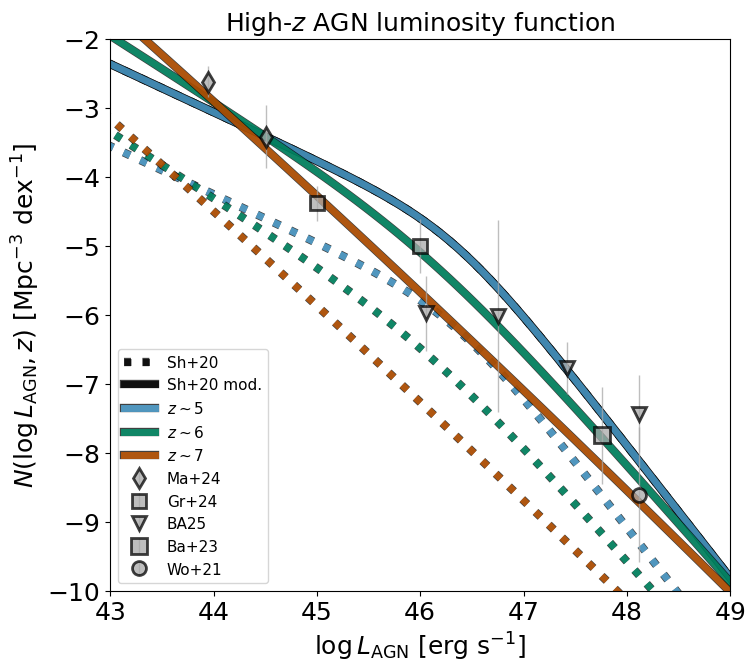}
    \caption{The AGN luminosity function at high redshift $z\gtrsim 4.5$. Colored lines refer to redshift $z\approx 5$ (blue), $6$ (green), and $7$ (red). Dotted lines represent the original luminosity function by \cite{Shen2020} used as our fiducial input in the main text, while the solid lines are a modified version (see Appendix \ref{app|AGN_LF}) to account for the recent data in the range $z\sim 5-7$ from \cite{Maiolino2024jades} (diamonds), \cite{Greene2024} (squares), \cite{Barlow2023} (squares), \cite{Barlow2025} (reversed triangles) and \cite{Wolf2021} (circle).}
    \label{fig|AGNLF_highz}
\end{figure}

The current situation may be visualized in Figure \ref{fig|AGNLF_highz}, where the AGN luminosity function by \cite{Shen2020} at redshift $z\sim 5-7$ (color-coded) is illustrated and compared to the most recent data collected by \cite{Barlow2025}. Besides the still large uncertainties in the data, there is a tendency for the fit by \cite{Shen2020} to somewhat under-predict the observational estimates. Remarkably, we checked that the rendition by \cite{Shen2020} can be reconciled with these new data just by modifying the parameter controlling the number density evolution $d_1\rightarrow d_1+0.2$ for $z\gtrsim 4.5$ (i.e., the negative number density evolution is slowed down). The results are illustrated by the solid lines, which agree with the data remarkably better.

One may wonder what would be the impact of this revised fit to the AGN luminosity function on the results presented in the main text. We have checked that this is irrelevant for the building up of the local BH mass function, since most of the accreted mass is acquired at redshift $z\lesssim 4$. In addition, the correction also turns out to be negligibly important for the GW background in the nano-Hz band, since most of it is originated at $z\lesssim 3$ (see Fig. \ref{fig|GWback}). A more quantitative analysis is presented in Appendix \ref{app|tests}.

\section{Additional tests}\label{app|tests}

In this Appendix we present additional tests on the robustness of the result on the expected GW background to variations in some assumptions of our analysis. We rely on the local BH mass function determination based on the stellar mass function and the $M_\bullet-M_\star$ relation, since this tends to maximize the impact of binary BH mergers. We focus on how the posterior on the GW normalized stretch amplitude is altered when the following assumptions are modified with respect to our fiducial analysis in the main text.

\begin{figure}[t!]
    \centering
    \includegraphics[width=1.\textwidth]{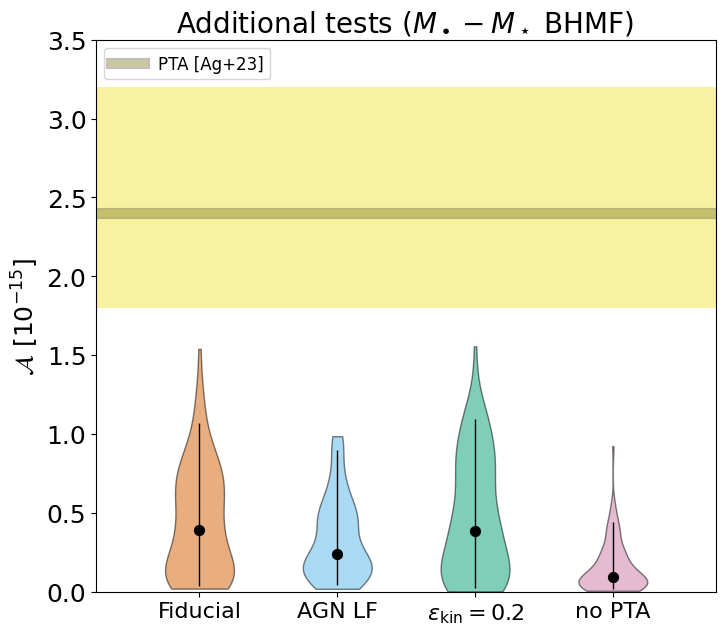}
    \caption{Additional tests on the contribution of binary BH mergers to the gravitational wave background. The local BH mass function determination from the stellar mass function and the $M_\star-M_\bullet$ relation is adopted. Colored violins illustrate the posterior distributions from our semi-empirical framework and MCMC analysis; the black dots and solid lines illustrate the median and the $2\sigma$ credible interval. The orange violin refers to our fiducial setting in the main text. The blue violin refers to the correction of the high-$z$ AGN luminosity function discussed in Appendix \ref{app|AGN_LF}. The green violin refers to a high value of the kinetic efficiency $\epsilon_{\rm kin}=0.2$. The magenta violin refers to the case when the PTA measurement of the GW background is not included in the likelihood. The measurement by the \texttt{NANOGrav} collaboration \cite{Agazie2023} is displayed by the yellow solid line with shaded area (median and $90\%$ credible interval).}
    \label{fig|tests}
\end{figure}

\begin{itemize}

\item AGN luminosity functions at high redshift. In the main text we have used the analytical rendition by \cite{Shen2020} over the redshift range $z\sim 0-7$. Here we try to employ the correction to the $z\gtrsim 4.5$ behavior discussed in Appendix \ref{app|AGN_LF} which is suggested by the most recent high$-z$ data.

\item Kinetic efficiency $\epsilon_{\rm kin}$. In the main text we have assumed $\epsilon_{\rm kin}=0.15\, \epsilon_{\rm thin}$ following the analysis by \cite{Shankar2008}; given the bestfit values of $\epsilon_{\rm thin}\lesssim 0.1$, rather small kinetic efficiencies are implied. However, since it has been claimed (e.g., \cite{Ghisellini2013,Zubovas2018}) that the kinetic efficiency could be larger, here we test the impact of assuming $\epsilon_{\rm kin}=0.2$. 


\item PTA measurement in the likelihood. In the main text we have included the GW background signal by \texttt{NANOGrav} as a proper measurement in the likelihood. Here we remove it from the likelihood, so that we just fit for the BH demography and clustering, and then check the GW background \emph{a-posteriori}.

The outcomes of these variations are illustrated in Fig. \ref{fig|tests}. Plainly, no major alteration with respect to the fiducial result of the main text is originated. In particular, adopting the corrected version of the high-$z$ luminosity function (which has an higher normalization) implies more mass added by accretion in the early buildup of the BH mass function, hence somewhat less room for mergers toward low redshifts where they become important. On the other hand, raising the kinetic efficiencies to values $\epsilon_{\rm kin}=0.2$ lowers somewhat the mass accumulated by accretion, and leave more room for mergers; however, the effect is  small since the relevant combination of parameter $(1-\epsilon-\epsilon_{\rm kin})/\epsilon$ depends mostly on the radiative efficiency $\epsilon$ than on the kinetic $\epsilon_{\rm kin}$. In addition, the change may be more effectively (in terms of the fit on the local BH mass function and clustering) compensated by varying the Eddington ratio distribution parameters than by raising the merger rates. Finally, eliminating the PTA measurements from the likelihood reduces the predicted background because the local BH mass function is already fitted well with the mass accumulated by accretion; thus the number of mergers is kept to values smaller than in the fiducial case, where instead the PTA measurement in the likelihood allows it to higher values, though still compatible with the BH demographics. 

\end{itemize}

\end{appendix}

\bibliographystyle{JHEP}
\bibliography{biblio}

\clearpage

\clearpage

\clearpage

\clearpage

\clearpage

\clearpage

\clearpage

\clearpage

\clearpage

\clearpage

\clearpage

\clearpage

\clearpage

\end{document}